\documentclass[aps,twocolumn,prd,showpacs,nofootinbib]{revtex4}

\usepackage{natbib}
\usepackage{amsmath}
\usepackage{graphicx}
\usepackage{subfigure}
\usepackage{bm}
\usepackage{amssymb}
\usepackage{latexsym}
\usepackage{graphicx}
\usepackage{color}
 \usepackage{ulem}

\newcommand{\sss}[1]{{\scriptscriptstyle{#1}}}
\newcommand{\dd}{\mathrm{d}}
\newcommand{\uPl}{\mathrm{Pl}}
\newcommand{\uin}{\mathrm{in}}
\newcommand{\uend}{\mathrm{end}}

\newcommand{\uc}{\mathrm{c}}
\newcommand{\usssPl}{\sss{\uPl}}
\newcommand{\ie}{\textsl{i.e.~}}

\newcommand{\eg}{\textsl{e.g.~}}

\newcommand{\apriori}{\textsl{a priori~}}
\newcommand{\aposteriori}{\textsl{a posteriori~}}
\newcommand{\ini}{\mathrm{in}}
\newcommand{\Mp}{M_\usssPl}
\newcommand{\nS}{n_{\mathrm{S}}}
\newcommand{\vev}{\textit{vev~}}
\newcommand{\vevs}{\textit{vevs~ }}

\def\be{\begin{equation}}
\def\ee{\end{equation}}
\def\bea{\begin{eqnarray}}
\def\eea{\end{eqnarray}}

\begin{document}
\title{Recursive Stochastic Effects in Valley Hybrid Inflation}

\author{Laurence Perreault Levasseur} \email{ l.perreault-levasseur@damtp.cam.ac.uk} \affiliation{DAMTP, Center for Mathematical Sciences, University of Cambridge \\ Cambridge CB3 0WA, United Kingdom}

\author{Vincent Vennin} \email{vennin@iap.fr} \affiliation{Institut d'Astrophysique de Paris, \\ UMR 7095-CNRS, Universit\'e Pierre et Marie Curie, \\ 98bis boulevard Arago, 75014 Paris, France}

\author{Robert Brandenberger} \email{rhb@hep.physics.mcgill.ca} \affiliation{Department of Physics, \\ McGill University, Montr\'eal, \\ QC, H3A 2T8, Canada}

\date{\today}

\begin{abstract}
 \indent Hybrid Inflation is a two-field model where inflation ends by a tachyonic instability, the duration of which is determined by stochastic effects and has important observational implications. Making use of the recursive approach to the stochastic formalism presented in Ref.~\cite{paper1}, these effects are consistently computed. Through an analysis of back-reaction, this method is shown to converge in the valley but points toward an (expected) instability in the waterfall. It is further shown that quasi-stationarity of the auxiliary field distribution breaks down in the case of a short-lived waterfall. It is found that the typical dispersion of the waterfall field at the critical point is then diminished, thus increasing the duration of the waterfall phase and jeopardizing the possibility of a short transition. Finally, it is found that stochastic effects worsen the blue tilt of the curvature perturbations by an $\mathcal{O}(1)$ factor when compared with the usual slow-roll contribution.
 
\end{abstract}

\maketitle

\section{Introduction}
\label{sec:intro}

Inflation is currently the leading paradigm attempting to shed light on the physics of the very early Universe. It describes a phase of accelerated expansion, which solves many problems of the hot big bang scenario~\cite{Brout:1977ix, Sato:1980yn, Kazanas:1980tx, Starobinsky:1980te, Guth:1980zm, Linde:1981mu, Albrecht:1982wi, Linde:1983gd}. It further provides a causal mechanism for generating fluctuations on cosmological scales, and predicts that their spectrum should be almost scale invariant, with small deviations from scale invariance which can be traced back to the precise microphysics of inflation~\cite{ Mukhanov:1981xt, Mukhanov:1982nu, Hawking:1982cz, Starobinsky:1982ee, Guth:1982ec, Bardeen:1983qw}. This prediction is consistent with the current astrophysical observation such as the CMB, including the measurement of the cosmic microwave background anisotropies. For this specific observable, the latest results \cite{Hou:2012xq, Sievers:2013ica, Ade:2013rta} give a slightly red tilted spectral index $\nS\simeq 0.96$, ruling out exact scale invariance $\nS=1$ at over $5\sigma$ and allowing to constrain the inflationary models still allowed by the data \cite{Martin:2013tda}.

With the ever-increasing precision of the experiments probing this window into the early Universe, it is now very important to develop robust and self-consistent methods for calculating inflationary predictions. For example, in the context of multi-field inflation, it is complicated to disentangle the gravitational and matter degrees of freedom when describing fluctuations produced in the  scalar fields using traditional methods. Typically, approximations are made to make the problem tractable which ignore back-reaction, that is, the effects of these very fluctuations on the background spacetime and fields trajectory. Restoring or even assessing the importance of these neglected effects then becomes extremely non-trivial, and it has been shown that such effects can have crucial impacts on the inflationary dynamics \cite{Tsamis:1992sx, Tsamis:1994ca, Mukhanov:1996ak, Abramo:1997hu} (see also Ref.~\cite{Brandenberger:2002sk} for a review of early work).

One way to re-sum these effects, at least partially, is to make use of the stochastic inflation formalism~\cite{Vilenkin:1983xp, Starobinsky:1986fx, Goncharov:1987ir, Nambu:1987ef, Nambu:1988je, Kandrup:1988sc, Nakao:1988yi, Nambu:1989uf, Mollerach:1990zf, Linde:1993xx, Starobinsky:1994bd}. The basic strategy is to derive an effective theory for the long-wavelength part of the fields, which are ``coarse grained'' at a scale larger than the Hubble radius. In this framework, the small-scale quantum fluctuations play the role of a ``bath'', and are collected in classical noise terms which affect the dynamics of the coarse-grained fields. The super-Hubble physics can thus be described by a stochastic classical theory.

The corresponding equations can be derived making use of the Schwinger-Keldysh closed time path formalism \cite{Keldysh:1964ud,Schwinger:1960qe,Jordan:1986ug,Calzetta:1986ey}, where the strategy is to split the degrees of freedom of the full quantum fields in momentum space through a window function, and perform the path integral over the small scale fluctuations. In Ref.~\cite{paper1}, this Lagrangian formulation of the theory is used to develop a recursive method for solving the stochastic equations when the background space-time is taken to be dynamic. It is this recursive method which we now wish to apply to models of multi-field inflation, and specifically to hybrid inflation. 

At the energy scale of inflation (typically around $10^{15}-10^{16}$ GeV), particle physics remains elusive, leaving room for a large variety of different inflationary scenarios. However, the supersymmetry- and supergravity- based extensions of the standard model of particle physics yield a well motivated model, hybrid inflation~\cite{ Linde:1993cn, Copeland:1994vg}, which can be realized in various ways in the context of $F$-term or $D$-term inflation~\cite{ Halyo:1996pp, Binetruy:1996xj, Dvali:1994ms, Kallosh:2003ux}. Hybrid Inflation is a two-field model where inflation is driven by a light inflaton field in a valley, where the potential is dominated by a vacuum constant, and where the inflaton \textit{vev} tunes the mass of an auxiliary field that becomes imaginary at some critical point, triggering the end of inflation by a ``waterfall'' phase. This auxiliary field is thus called ``waterfall field''. This model is known to lead, in the valley, to a blue spectrum $\nS>1$ for the cosmological fluctuations, a prediction which is strongly disfavored by current observations.

However, it was shown~\cite{Clesse:2008pf, Abolhasani:2010kn} that, in some regions of parameter space, a significant number of $e$-folds can occur in the waterfall regime. In this case, it was also demonstrated that the spectral index becomes red, eliminating this tension. Since the duration of the waterfall phase is determined by the stochastic dispersion of the waterfall field at the critical point, it is therefore crucial to compute this quantity accurately, and to properly include the stochastic effects in the description of the model.

Moreover, in Ref.~\cite{Martin:2011ib}, it was shown that stochastic effects can significantly alter the inflationary background dynamics in the context of hybrid inflation, especially close to the critical point where the two-field potential is very flat and where one enters a regime of stochastically-driven saddle-point inflation. Back-reaction is therefore expected to be important there, and the associated mode coupling effects were investigated in Ref.~\cite{Levasseur:2010rk}.

In the current paper, these issues are addressed by applying the new recursive method developed in Ref.~\cite{paper1} to the specific example of two-field hybrid inflation potentials, to illustrate how this formalism can be implemented, and how it yields new results when compared with standard techniques for computing observables of inflation beyond the leading order. The outline of the strategy is to first calculate the background evolution in the presence of a free noise, then the corrected quantum noise on this shifted background, to finally come back to the background, coarse-grained dynamics shifted in light of the fluctuations, and so on until the process converges. We demonstrate the convergence of this method in the valley region, where the usual QFT methods of perturbation theory are known to be under control. Doing so, we calculate modified predictions emerging from a consistently implemented non-perturbative method for cosmological observables such as the tilt of the CMB power spectrum. Most interestingly, we identify regimes of hybrid inflation where stochastic effects dominate over regular perturbative corrections.

One of the main interests of this program of research is however the waterfall phase, where back-reaction and mode coupling effects are expected to be important. We discuss important implications of the findings of the current work for this phase, but plan to pursue a more thorough study of the waterfall phase in the future.

The paper is organized as follows. In section \ref{sec:ClassAndModes}, the background classical dynamics of valley hybrid inflation is computed, and the recursive approach to stochastic inflation of Ref.~\cite{paper1} is presented. A first-step massless de-Sitter solution, valid up to $\mathcal{O}(\hbar)$ and to zeroth order in slow-roll, is presented.

In section \ref{sec:PertExpAndModeCalc}, we move on to calculate the value of the noises up to $\mathcal{O}(\hbar^2)$ and to leading order in slow-roll. To do so, we make use of the fact that solving for the propagators of the bath fields at this order is equivalent to solving the linearized quantum mode functions in a shifted background. We compute the amplitude of these linear perturbations in both fields, and identify different regimes for the waterfall field fluctuations evolution. 

In section \ref{sec:sigmapsi}, the corresponding modified amplitudes of the noise terms are implemented in the stochastic equations. Their effect on the mean deviation in the waterfall direction is carefully computed. Short-lived waterfalls are shown to be unlikely, since the quasi-stationary time behavior of the auxiliary field distribution breaks down in this regime, reducing its quantum dispersion at the critical point, hence lengthening this stage. Furthermore, an analysis of back-reaction show that the recursive process converges in the valley but blows up in the waterfall, suggesting perturbative instability there. 

In section \ref{sec:influcts} we study how the classical inflation perturbations beyond zeroth order in the slow-roll expansion are influenced by stochastic effects, in particular when it comes to the curvature perturbations spectral tilt. We obtain that the stochastic effects worsen the blue tilt problem, by a factor $\mathcal{O}(1)$ compared to the usual slow-roll contribution. Finally in section \ref{sec:conclusion}, we summarize our main findings and suggest possible further investigations.

\section{Valley Hybrid Inflation}

\label{sec:ClassAndModes}

The potential of hybrid inflation in the field space $\left(\Phi,\Psi\right)$, where $\Phi$ is the inflaton and $\Psi$ the waterfall field, is given by:
\be 
\label{hybridInflationpotential}
	V(\Phi,\Psi)=\frac{1}{2}m^2\Phi^2+\frac{\lambda}{4}(\Psi^2-v^2)^2+\frac{g^2}{2}\Phi^2\Psi^2 \, .
\ee
The true minima of the potential are located at $\Phi=0$ and
$\Psi=\pm v$, while the instability point is given by
\be
	\Phi^2_\uc=\frac{v^2\lambda}{g^2}\,, \quad\Psi_\uc=0\,. 
\ee
It is usually assumed that hybrid inflation occurs in the vacuum dominated regime, for which $\Phi_\uc<\Phi\ll\lambda^{1/2}v^2/m$ and $\Psi\ll v$. In this approximation, the first slow-roll parameter in the valley ($\Phi>\Phi_\uc$, $\Psi=0$) is given by $\varepsilon_1\simeq 8 m^4 \phi^2 \Mp^2/(\lambda^2 v^8) $, hence for the slow-roll approximation to be satisfied in the valley, one must assume that $\lambda v^4 \gg m^2\Phi_\uc\Mp$, $\Mp$ being the reduced Planck mass. In the same manner, the smallness of the second slow-roll paramter $\varepsilon_2\simeq -8\Mp^2m^2/(\lambda v^4)\ll 1$ implies the more stringent condition $\lambda v^4 \gg m^2\Mp^2$ (one then has $\varepsilon_1\ll\varepsilon_2$). In this case, the total energy density is dominated by the constant term of the potential $\rho\simeq V\simeq\lambda v^ 4/4$. Motivated by the supersymmetric version of the model, we also take $\Phi_\uc \simeq v$, or, using the definition of $\Phi_\uc$ in terms of the potential parameters, $\lambda^{1/2}\simeq g$. Finally, in order for the model to be consistently derived, inflation must proceed at small values of the fields (compared to the Planck mass), and one can consider that $\Phi_\uc,v\ll \Mp$. The constraints on the potential parameters coming from these considerations and the ones following below are collected in appendix~\ref{app:NotationsAndAssumptions}, together with a summary of the notations used throughout the paper.

Taken literally, this model produces a blue tilt for the spectrum of cosmological perturbations \cite{Lyth:1998xn} when inflation is realized in the valley,
\be 
n_{\mathrm{s}}\simeq 1-\varepsilon_2\simeq1+8\frac{\Mp^2m^2}{\lambda v^4}\, .
\ee
Recently, to alleviate this problem, it has been suggested to realize the last $60$ $e$-folds of inflation in the waterfall phase \cite{Clesse:2010iz}. In order to do so, one must choose the parameters of the potential in order for a sufficiently large number of $e$-folds to be realized in the waterfall phase, making the model behaving in a fashion effectively similar to a (multi-field) hill-top model. The duration of this stage being determined by the mean stochastic shift of the waterfall field at the critical point, an accurate calculation of the preceding stochastic effects in the valley is crucial to determine whether such a scenario is viable or not.

Note that one could also choose to glue a different potential for the inflaton in the valley phase, chosen specifically in order to produce the desired tilt, and then use the symmetry breaking shape of the hybrid potential for the sole purpose of ending inflation (see, e.g. Refs.~\cite{Lyth:1996we,Lyth:1996kt,King:1997ig}). However, as we will see, when choosing the potential, one has to be careful that stochastic effects do no re-introduce the blue tilt. In any case, $m^2 \phi^2$ is the simplest choice for the inflaton potential, and in the absence of special symmetries (e.g. conformal symmetry) such a term will be present and will dominate at small field values. Thus, as a toy model for multi-field inflation, the terms included in our potential are the lowest order terms one would expect to find.

\subsection{Classical Dynamics}
\label{subsec:classical}
\par

In this subsection, we study the classical behavior of the inflaton and waterfall fields at the background level, which represents the first step of the recursive method presented below. The slow-roll equations controlling the evolution of the classical background fields $\varphi^{\left(0\right)}$ and $\chi^{\left(0\right)}$  can be expressed as
\begin{eqnarray}
\label{KGphi}
3H^2\frac{\mathrm{d}\varphi^{\left(0\right)}}{\mathrm{d}N} &\simeq& 
-m^2 \varphi^{\left(0\right)}\left(1+\frac{g^2 {\chi^{(0)}}^2 }
{m^2}\right)\, ,\\
\label{KGpsi}
3H^2\frac{\mathrm{d}\chi^{\left(0\right)}}{\mathrm{d}N} 
&\simeq& 
-\lambda v^2{\chi^{(0)}}\left(\frac{{\varphi^{\left(0\right)}}^2
-\Phi_\mathrm{c}^2}{\Phi_\mathrm{c}^2}
+\frac{{\chi^{\left(0\right)}}^2}{v^2}\right)\, , 
\end{eqnarray}
with
\be
H^2 = \frac{1}{3\Mp^2} \rho \simeq\frac{\lambda v^4}{12\Mp^2}\, .
\ee
The superscript $^{(0)}$ denotes a background, homogeneous quantity and $H=\dot{a}/a$ is the Hubble parameter with a dot standing for a derivative with respect to cosmic time $t$. The quantity $N$ is the number of $e$-folds, $N\equiv \ln(a/a_\mathrm{i})$, where\footnote{The cosmological scale factor is denoted by $a(t)$.} $a_\mathrm{i}$ is the scale factor at an initial reference point.

If inflation starts beyond the critical line $\Phi=\Phi_\uc$, the system very quickly reaches the region where $\chi^{(0)}\ll v$ and inflation is driven by the inflaton $\varphi^{\left(0\right)}$ which slowly rolls down towards the critical point, while the waterfall field $\chi^{\left(0\right)}$ first undergoes damped oscillations at the bottom of the valley, before experiencing a short simple damping regime. Defining
\be 
\omega\left(N\right)=\frac{3}{2}\sqrt{1-\frac{16}{3}\frac{\Mp^2}{v^2}\frac{\frac{{\varphi^{(0)}}^2\left(N\right)}{\Phi_\uc^2}-1}{1+\frac{2m^2}{\lambda v^4}{\varphi^{(0)}}^2\left(N\right)}}\, ,
\ee
the homogeneous time evolution of these two fields is given by
\begin{equation}
\label{phi0trajValley}
\varphi^{\left(0\right)}=\varphi_\mathrm{in}\exp
\left(-4\frac{\Mp^2m^2}{\lambda v^4}
N\right)\ ,
\end{equation}
\begin{equation}
\label{psi0trajValley}
\chi^{\left(0\right)}=\left\lbrace
\begin{array}{c l}  
&\chi_\uin\frac{{\rm e}^{-3N/2}}{\sqrt{2\omega (N)}}
\left[C_1{\rm e}^{I\left(N\right)}
+C_2{\rm e}^{-I\left(N\right)}\right]
\nonumber\\ &\quad \quad \quad \quad \quad \quad \quad \quad \quad
\mathrm{if}\ \frac{\varphi^{\left(0\right)}}{\Phi_\uc}>1
+\frac{3}{32}\frac{v^2}{\Mp^2}\, ,
\nonumber\\
&\chi_\uin\exp\left[-4\frac{\Mp^2}{M^2}
\left(\frac{\varphi_\uin}{\Phi_\uc}-1\right)N\right]
\nonumber\\ &\quad \quad \quad \quad \quad \quad \quad \quad \quad
\mathrm{if}\ \frac{\varphi^{(0)}}{\Phi_\uc}<1+\frac{3}{32}
\frac{v^2}{\Mp^2}\, ,
\end{array}
\right.
\end{equation}
where $\varphi_\uin$ and $\chi_\uin$ are the initial inflaton and waterfall values, $C_1$ and $C_2$ are integration constants, and where $I\left(N\right)$ is given by
\begin{eqnarray}
I(N)&\simeq & -\frac{\sqrt{3}}{\Mp}\frac{\lambda v^3}{2m^2}
\Biggl[\sqrt{\frac{{\varphi^{\left(0\right)}}^2}
{\Phi_\mathrm{c}^2}-1}-\arctan \left(\sqrt{\frac{{\varphi^{\left(0\right)}}}
{\Phi_\mathrm{c}^2}-1}\right)
\nonumber \\ 
& &-\sqrt{\frac{\varphi_\ini^2}{\Phi_\mathrm{c}^2}-1}
+\arctan \left(\sqrt{\frac{\varphi^2_\ini}
{\Phi_\mathrm{c}^2}-1}\right)
\Biggr].
\end{eqnarray}
From the previous equations, the total number of $e$-folds spent in the valley is
\begin{equation}
\label{Nc}
N_\uc=\frac{\lambda v^4}{4m^2\Mp^2}\ln\left(\frac{\varphi_\uin}{\Phi_\uc}\right)\, .
\end{equation} 
It is typically a large number because of our assumption $\lambda v^4/(2m^2)\gg \Mp^2$. Finally, the value of $\chi$ at the end of this stage reads
\begin{equation}
\label{psic}
\chi_\uc^{(0)}=\chi_\uin\exp\left[-2\frac{\lambda v^ 2}{m^ 2}
\left(\frac{\varphi_\uin}{\Phi_\uc}-1\right)
\ln\left(\frac{\varphi_\uin}{\Phi_\uc}\right)\right]\, .
\end{equation}
With the assumptions made above on the potential parameters, this value is typically so small that it is completely washed by the quantum noise that we calculate in the rest of the paper. The number of $e$-folds spent during the waterfall phase is given by \cite{Kodama:2011vs,Martin:2011ib}  
\begin{equation}
\label{eq:efoldswater}
N_\uend-N_\mathrm{c}\simeq \frac{\lambda^{1/2}v^3}{4m\Mp^2}
\ln ^{1/2}\left(\frac{m}{g \chi_\mathrm{c}}\right)\, .
\end{equation}
From this, if one is interested in the regime where the required $\sim 60$ $e$-folds of inflation take place during the waterfall phase, one needs to work with $\lambda^{1/2}v^3 \gg m\Mp^2$. Note that a more detailed description of the waterfall phase is reviewed in appendix~\ref{app:waterfall}.

Finally, inflation stops when $\varepsilon_1=1$ and the system starts oscillating around one of the two true minima of the potential, triggering a phase of (p)reheating \cite{GarciaBellido:1997wm,GarciaBellido:1998gm,Copeland:2002ku,Finelli:2000ya}.

\subsection{Stochastic Formalism and Recursive Strategy}
\label{sec:RecursiveStrategy}

The previous subsection details the dynamics of two classical fields $\varphi^{\left(0\right)}$ and $\chi^{\left(0\right)}$, each obeying a homogeneous Klein-Gordon equation. The system we are interested in studying, however, is a system consisting in two inhomogeneous four-dimensional quantum fields, $\Phi$ and $\Psi$. Solving the full Heisenberg field equations they obey in curved spacetime is in general impossible with current techniques, and so different approximation schemes are typically applied to make the calculations tractable.

One such strategy is to derive an effective theory for the classicalized, long wavelength part of the fields, which can be shown \cite{Kiefer:1998qe,Polarski:1995jg} to behave as a classical stochastic system. The super-Hubble Fourier modes of the full quantum fields, corresponding to scales with $k<\epsilon a H$ ($\epsilon < 1$ being a small dimensionless parameter setting the averaging scale and collecting only the super-Hubble sufficiently-squeezed and decohered modes) are collected into ``coarse-grained'' fields $\varphi$ and $\chi$. These coarse-grained fields evolve in the presence of a quantum bath made of the remaining, sub-Hubble modes with $k>\epsilon a H$, which are collected using a window function $W_H(k,t)$ in the Fourier expansion of the corresponding full quantum fields. The fields of the quantum bath and the coarse grained fields are thus given by
\bea
\phi_> &=& \int\frac{\dd ^3k}{(2\pi)^3}W_H\left(k, t\right)\left[\phi_{\bf k}\hat{a}_{\bf k}e^{-i {\bf x\cdot k}}+\phi^*_{\bf k}\hat{a}^\dagger_{\bf k}e^{i {\bf x\cdot k}} \right]
\label{modefunctiondefphi}
\, ,\nonumber\\ & & \\
\varphi &\equiv& \Phi-\phi_>\label{phibathsystsplit}\nonumber\, ,\\
\psi_> &=& \int\frac{\dd ^3k}{(2\pi)^3}W_H\left(k, t\right)\left[\psi_{\bf k}\hat{b}_{\bf k}e^{-i {\bf x\cdot k}}+\psi^*_{\bf k}\hat{b}^\dagger_{\bf k}e^{i {\bf x\cdot k}} \right]
\label{modefunctiondefpsi}
\, ,\nonumber\\ & & \\
\chi &\equiv& \Psi-\psi_>\label{psibathsystsplit}\nonumber\, ,
\eea
where the quantum bath fields have been written in terms of their linearized mode functions $\phi_{\bf k}$ and $\psi_{\bf k}$, and the creation and annihilation operators $\hat{a}^\dagger_{\bf k}, \hat{a}_{\bf k}, \hat{b}^\dagger_{\bf k}, \hat{b}_{\bf k}$.

Because Fourier modes constantly cross the Hubble radius during inflation, leaving the quantum bath to join the coarse-grained fields, the quantum bath sources the coarse-grained part of the fields. This effect adds to their equations of motion a stochastic noise term, yielding, to leading order\footnote{We work in units where $\hbar=1$ all along the paper, and will not write explicitly the factors of $\hbar$ to avoid making the notation heavier. However, the power counting from the expansion in $\hbar$ should be straightforward to restore from the text.} in $\hbar$:  
\begin{eqnarray}
\label{Langphi}
3H^2\frac{\mathrm{d}\varphi}{\mathrm{d}N} &=& 
-m^2 \varphi\left(1+\frac{g^2 \chi^2 }
{m^2}\right)
+3H\xi_\phi\left(N\right)\, ,\\
\label{Langpsi}
3H^2\frac{\mathrm{d}\chi}{\mathrm{d}N} &=& 
-\lambda v^2
\chi\left(\frac{\varphi^2-\Phi_\mathrm{c}^2}{\Phi_\mathrm{c}^2}
+\frac{\chi^2}{v^2}\right)
+3H\xi_\psi\left(N\right)\, ,\nonumber  \\
\end{eqnarray}
where $\xi_\phi$ and $\xi_\psi$ are two uncorrelated white Gaussian noises with zero mean and variance given by
\begin{eqnarray}
\left\langle \xi_\phi\left(N\right)\xi_\phi\left(N^\prime\right) 
\right\rangle =\nonumber
\quad\quad\quad & \quad\quad\quad
\\  \quad\quad
\frac{\epsilon^3H^5}{2\pi^2}a^3 
\left\vert {\phi}_{\bm{k}}\right\vert_{k=\epsilon a H}^2& \left(1-2 
\varepsilon_1 \right) \delta \left(N-N^\prime\right)\, , & 
\label{noisecorrphi} \\
\left\langle \xi_\psi\left(N\right)\xi_\psi\left(N^\prime\right) 
\right\rangle =\nonumber
\quad\quad\quad & \quad\quad\quad
\\  \quad\quad
\frac{\epsilon^3H^5}{2\pi^2}a^3 
\left\vert {\psi}_{\bm{k}}\right\vert_{k=\epsilon a H}^2& 
\left(1-2\varepsilon_1\right)\delta\left(N-N^\prime\right)\, . &
\label{noisecorrpsi}
\end{eqnarray}
Here, ${\psi}_{\bm{k}}$ and ${\phi}_{\bm{k}}$ are evaluated at the time when they join the coarse grained scales.

In principle other noise terms arise in Eqs.~(\ref{Langphi}-\ref{Langpsi}), which are either suppressed by factors of $\varepsilon_1\ll 1$, or come from higher order contributions of the loop expansion in the quantum piece of the fields, and are therefore suppressed by higher ({\it i.e.} at least two) powers of $\hbar$. The latter contribution mainly implements mode coupling effects, which are not taken into account in the current paper. 

Notice that the time variable used in these equations is the number of $e$-folds $N$, since it was shown in Refs.~\cite{Finelli:2008zg,Finelli:2010sh} that this time gauge must be used to decribe the stochastic dynamics of the gauge-invariant Mukhanov variables for the fields. Other choices of time variable would in principle correspond to different stochastic processes. However, note that, under the vacuum-domination approximation (which we are assuming here), the mapping between number of $e$-folds, cosmic time and conformal time is non stochastic and one can equivalently work in different time gauges. 

These equations (\ref{Langphi}) and (\ref{Langpsi}) are Langevin equations which describe Markovian processes, which means that instead of having to solve a single equation of motion (single ``realization''), one now has to calculate a whole probability distribution $\rho\left(\varphi,\chi,N\right)$ over many realizations, through a Fokker-Planck equation. Expectation values of functionals of the stochastic fields, in particular their correlation functions, can be calculated by averaging over realizations using $\rho$ as an integral kernel~\cite{Starobinsky:1994bd}.

One can see that the noise amplitudes appearing in  Eqs.~(\ref{noisecorrphi}-\ref{noisecorrpsi}) at a given time $N$ are computed from the amplitudes of the linearized Fourier modes of the bath fields crossing the $\epsilon$-scaled Hubble radius at time $N$. This simple form holds only if one chooses the window function entering the definition of the quantum fields to be a Heaviside step-function, with a transition at $k=\epsilon a H$ (for a discussion of the influence of the choice of the window function, see \eg Refs.~\cite{Winitzki:1999ve, Casini:1998wr}). Regardless of the choice of window function, one needs to solve the linearized Fourier mode function equations for each fields and evaluate the solutions, $\phi_{\bm{k}}$ and $\psi_{\bm{k}}$, at Hubble crossing in order to obtain the noise amplitudes.

We therefore obtain a system of two coupled sets of equations: on one hand, the set of Langevin equations for the two stochastic processes $\varphi$ and $\chi$, and, on the other, the set of linearized mode function equations for  $\phi_{\bm{k}}$ and $\psi_{\bm{k}}$. This is where the recursive strategy of Ref.~\cite{paper1} comes into play. Let us see how it proceeds. One should keep in mind that the Langevin equations~(\ref{Langphi}-\ref{Langpsi}) arise from a Lagrangian theory, in which the small wavelength fluctuations are integrated out to yield an effective theory for the coarse-grained field. Such fluctuations are evolved by equations of motion that involve coarse-grained - or ``background'' - quantities, the dynamics of which is itself shifted by these small wavelength fluctuations. This forms a closed system of equations that is in general very difficult to solve. Indeed, at each time $N$, one needs to compute the amplitude of the modes that are crossing the Hubble radius, which depends on the previous history of the background, which is itself determined by the amplitude of all the modes that previously crossed the Hubble radius.

Considered as a whole in this manner, the process stops to be Markovian since the amplitudes of the noise at a given time $N$ depend on all the realizations of the noises at previous times $N^\prime<N$, and one needs to assign a so called ``prescription'' $\alpha \in [0,1]$ to the Langevin equations (which sets at which point $N + \alpha \mathrm{d} N$ the noises must be calculated when the fields are incremented between $N$ and $N+ \mathrm{d}N$, when defining the Langevin dynamics as a limit of a discrete stochastic process). The resulting integro-differential equation becomes in practice impossible to solve.

However, as argued in Ref.~\cite{paper1}, a perturbative solution can be obtained by recursively solving a sequence of Markovian processes. To this end, one first evolves the linearized Fourier mode functions for each field to zeroth order in the slow-roll parameters and to first order in $\hbar$. This means evolving the mode equations truncated as if they were massless equations over exact de Sitter space. This enables one to calculate the (zeroth order) noise amplitudes at each time $N$, and to obtain the corresponding driving term at every time in Eqs.~(\ref{Langphi}-\ref{Langpsi}), giving us the leading $\hbar$ quantum effects to the coarse-grained equations. Solving the latter now keeping only terms to zeroth order in slow-roll and leading order in $\hbar$ then provides one with the shifted (or renormalized) associated background fields.

One can then solve again the equations of motion for the linearized mode functions of the quantum fields, this time in the presence of a ``mean'' background calculated from averaging over many realizations of the coarse-grained system described by the Langevin equations at this order (or using the pdf obtained by solving the Fokker-Planck equation). This enables one to calculate new noise amplitudes which include corrections of leading order in slow-roll and second order in $\hbar$ (note, however, that at this point one cannot yet make predictions about the classical spectrum of perturbations). 

From these noise amplitudes valid to higher order, one can go back to the Langevin system for the coarse-grained fields,  Eqs.~(\ref{Langphi}-\ref{Langpsi}), and find new, corrected solutions. These will now be valid up to next-to-leading order in $\hbar$ and to leading order in slow-roll parameters
. From these corrected solutions to the classical system, one can study classical perturbations of the coarse-grained fields and make predictions beyond zeroth order in slow-roll, for example, for the spectral index. 

One can keep solving recursively the linearized mode functions (describing the quantum bath and required to calculate the noise amplitudes) and the Langevin equations (describing the coarse-grained classical fields) until one reaches the required level of accuracy. If such a process converges towards a limit point, it should be close to the actual solution of the implicit closed equations. If, on the contrary, it does not possess any fixed point, this should be interpreted as the sign that the back-reaction effects may be out of control and that the whole model is under pressure. In any case, performing such a program is of interest and we now carry it out for the model being considered in this paper.

\subsection{Coarse-grained system up to zeroth order: Massless de-Sitter solution}
\label{sec:masslessDSdispersions}

As a first step, let us assume that the linearized mode functions for the bath fields, $\phi_{\bm{k}}$ and $\psi_{\bm{k}}$ with $k>\epsilon aH$, are free and massless and evolving in a de-Sitter background. Since the potential is vacuum dominated in the valley phase, the de-Sitter approximation seems to be well justified. The inflaton perturbations $\phi_{\bm{k}}$ also need to be very light with $m\ll H$ in order for slow-roll inflation to proceed, as already mentioned. However, the waterfall perturbations $\psi_{\bm{k}}$ can \apriori be very massive (it is precisely the mass of the waterfall field that quickly brings the system to the bottom of the valley), and thus the approximation of masslessness may be totally unjustified for this field. This is the object of the calculation and discussion of section~\ref{sec:sigmapsi}. For now, we will assume that since inflation proceeds as $\Phi$ approaches the critical point, and $\Psi$ becomes lighter and lighter, the approximation correspondingly becomes better and better, so that close enough to the critical point, the following calculation is a reliable first step  result.

The standard massless de-Sitter solution gives
\be
\left|\phi_{{\bf k}} \right|^2_{k=\epsilon aH}=\left|\psi_{{\bf k}} 
\right|^2_{k=\epsilon aH}= \frac{H^2}{2(\epsilon a H)^3},
\ee
so that,  to leading order, we obtain the correlators:
\begin{eqnarray}
\label{noisecorrphideSitter}
\left\langle \xi_\phi\left(N\right)\xi_\phi\left(N^\prime\right) 
\right\rangle =\frac{H^4}{4\pi^2}\delta\left(N-N^\prime\right)\, , 
\\
\left\langle \xi_\psi\left(N\right)\xi_\psi\left(N^\prime\right) 
\right\rangle =\frac{H^4}{4\pi^2}\delta\left(N-N^\prime\right)\, , 
\label{noisecorrpsideSitter}
\end{eqnarray}
hence the well known $H/2\pi$ noise amplitude commonly used in stochastic inflation. Let us now try to assess the typical dispersion acquired by the field distributions when subjected to the influence of these stochastic effects.

\begin{figure*}
\begin{center}
\includegraphics[width=8.5cm]{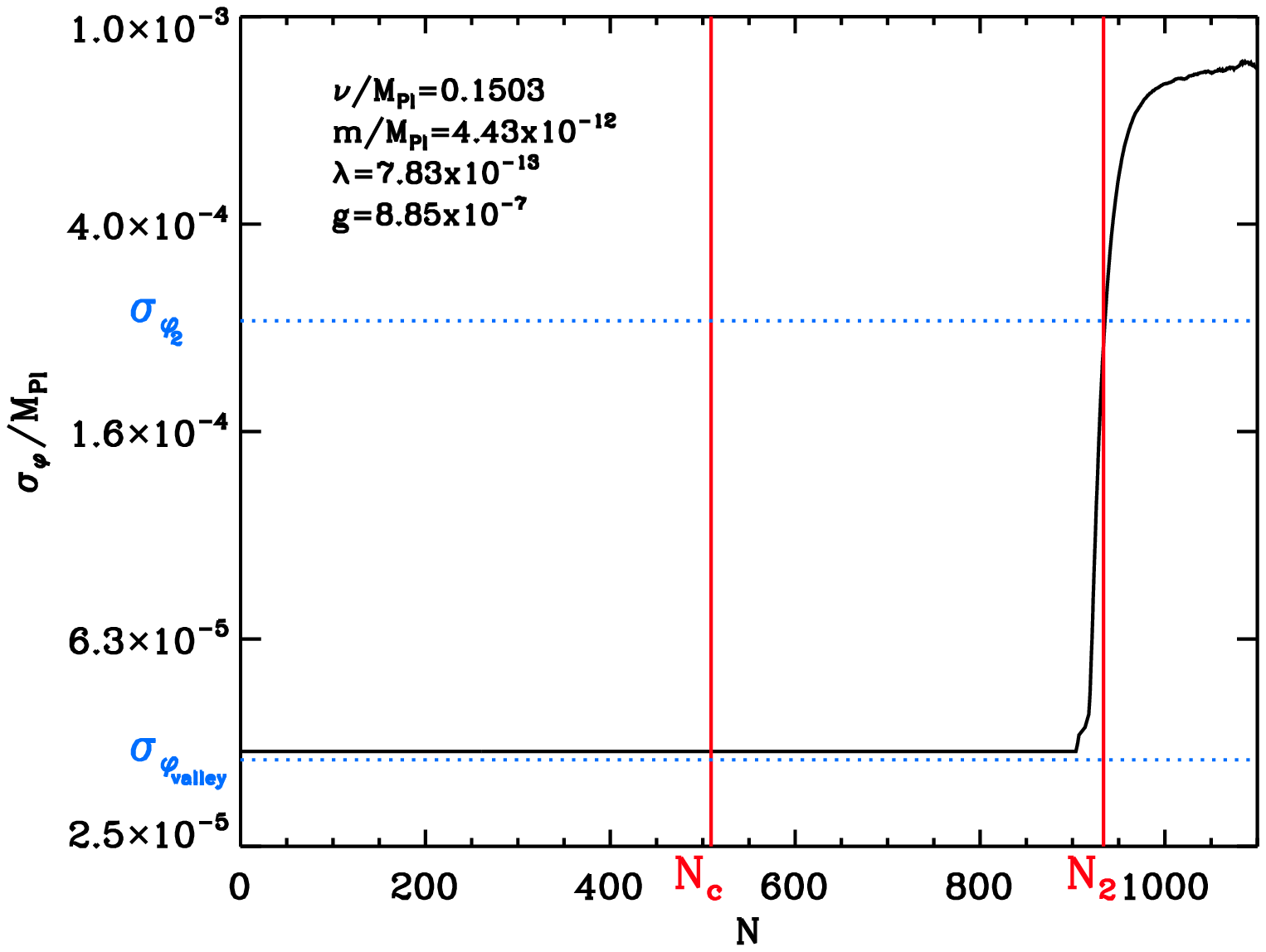}
\includegraphics[width=8.5cm]{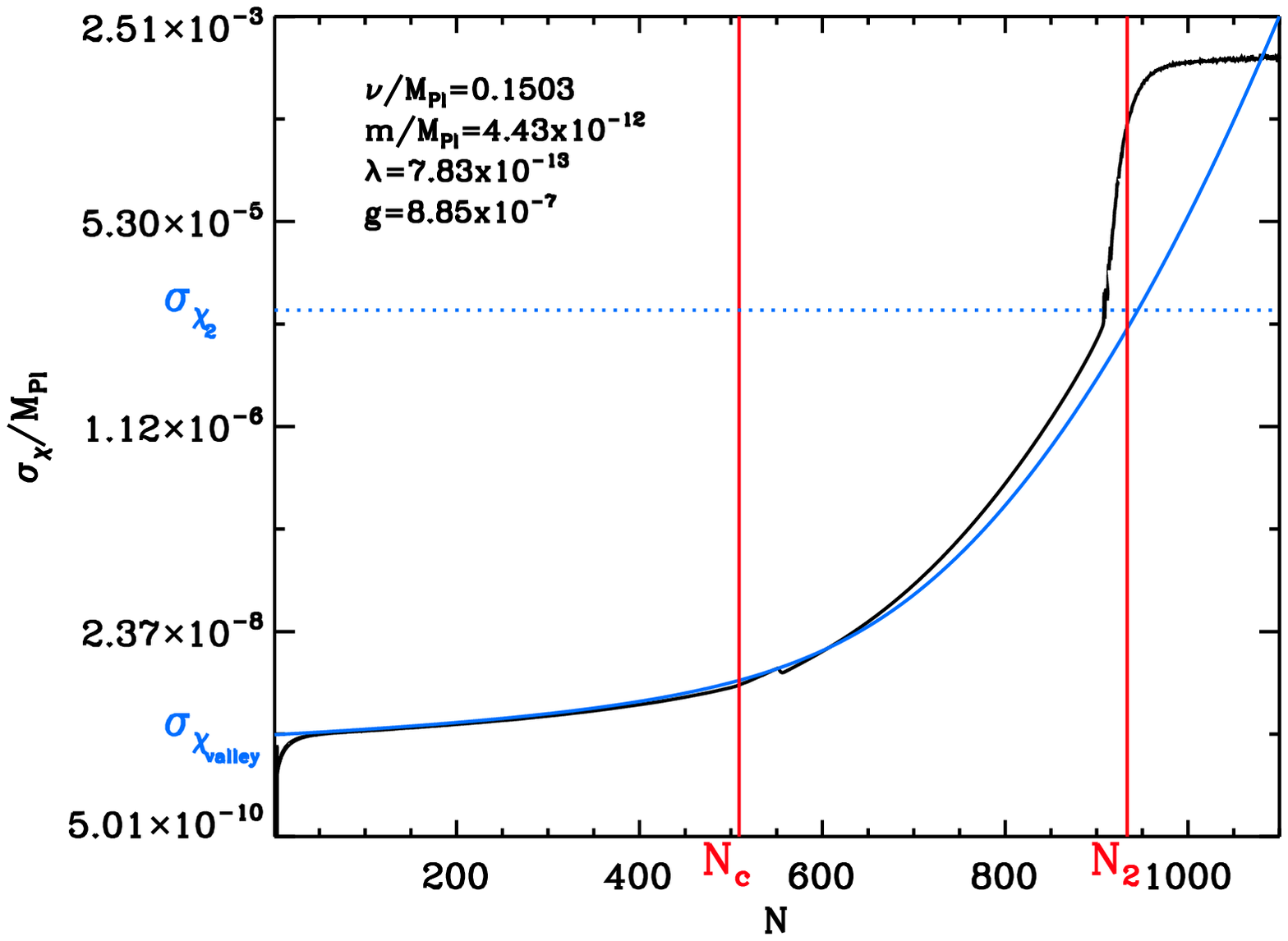}
\caption{Dispersion of the inflaton field (left panel) and of the 
waterfall field (right panel) during inflation. The values of $N_\uc$ and $N_2$ are given by Eq.~(\ref{Nc}) and Eq.~(\ref{eq:N2}), respectively. The constant dispersion in the valley $\sigma_{\varphi_\mathrm{valley}}$ (blue dotted line) and the time-dependent dispersion $\sigma_{\chi_\mathrm{valley}}$ (blue solid line) correspond to the values given by Eq.~(\ref{sigmaphi}) and Eq.~(\ref{sigmapsi}), respectively, while the dispersion in the $\varphi$ direction $\sigma_{\varphi_2}$ and in the $\chi$ direction $\sigma_{\chi_2}$ (blue dotted lines) at the end of the sub-phase 2 of the waterfall phase correspond to the value given by Eq.~(\ref{sigmaphi2}) and Eq.~(\ref{sigmapsi2}), respectively. The black lines correspond to numerical results coming from solving $\sim 10^7$ realizations of the Langevin equations.}
\label{fig:dispersions}
\end{center}
\end{figure*}

The importance of the stochastic effects in the $\Phi$ direction can be estimated through the ratio, which we call $\Delta_\Phi$, of the mean magnitude of the quantum kick $H/(2\pi)$ during a typical time interval of one $e$-fold, to the typical classical change in the inflaton value $\simeq M^2\left(\partial V/\partial\varphi\right)/V$ during the same time interval. In the valley close to the critical point, one obtains
\begin{equation}
\Delta_\Phi=\frac{1}{16\sqrt{3}\pi}\lambda g\frac{v^5}{m^2\Mp^3}\, .
\end{equation}
If $\Delta_\Phi\ll 1$, the inflaton dynamics in the valley phase is dominated by its classical drift so that the classical solution~(\ref{phi0trajValley}) can be used in Eq.~(\ref{Langpsi}). We will restrict our attention to this case. In particular, we will not consider the eternal version of hybrid inflation which is obtained if the mass is chosen to be so small that $\Delta_\phi \gg 1$. Letting
\begin{equation}
x\equiv \mathrm{e}^{-8\frac{\Mp^2m^2}{\lambda v^4} \left(N-N_\uc\right)}\, ,
\end{equation}
the $\chi$ equation of motion (\ref{Langpsi}) can be rewritten as:
\begin{equation}
\frac{\dd \chi}{\dd x}=\frac{\lambda v^2}{2m^2}\frac{x-1}{x}\chi 
-\sqrt{\frac{3}{2x}}\frac{\xi_\psi(x)}{m}\, ,
\end{equation}
where $\xi_\psi(x)$ shares the same statistical properties (\ref{noisecorrpsi}) as $\xi_\psi(N)$, replacing $N$ by $x$ in the delta function argument. The solution to this equation is given by:
\begin{eqnarray}
\label{eq:chistoSol}
\chi\left(x\right)&=& 
C\exp\left[\frac{\lambda v^2}{2m^2}\left(x-\ln x\right)\right]
\nonumber\\& &
-\sqrt{\frac{3}{2}}
\exp\left[\frac{\lambda v^2}{2m^2}\left(x-\ln x \right)\right] 
\times\nonumber\\& &
\int_\infty^{x}\exp\left[-\frac{\lambda v^2}{2m^2}
\left(x^\prime-\ln x^\prime \right)\right] 
\frac{\xi_\psi\left(x^\prime\right)}{m}\frac{\dd x^\prime}
{\sqrt{x^\prime}}
\, ,\nonumber\\
\end{eqnarray}
where $C$ is a constant of integration. It is set to $C=0$ provided one assumes an initial delta distribution for $\chi$ at $\varphi\gg\Phi_\uc$ (\ie $x\rightarrow \infty$). In this case, using Eq.~(\ref{noisecorrpsideSitter}), the two-point correlation function can be calculated to be:
\begin{equation}
\label{twopointscorrfctPsi}
\left\langle\chi^2\right\rangle=\frac{1}{384\pi^2}
\frac{\lambda^2v^8}{m^2\Mp^4} 
\left(\frac{m^2\mathrm{e}^x}{\lambda v^2 x}\right)^
{\frac{\lambda v^2}{m^2}} 
\Gamma\left(\frac{\lambda v^2}{m^2},\frac{\lambda v^2}
{m^2}x\right)\, ,
\end{equation}
where $\Gamma$ is the upper incomplete gamma function. Therefore, the dispersion of the distribution for $\chi$ is found to be:
\begin{eqnarray}
\label{sigmapsi}
\sigma_\chi&\equiv&\sqrt{\left\langle\chi^2\right\rangle-\left\langle\chi\right\rangle^2}
\nonumber\\&=&
\frac{\lambda v^4}{8\sqrt{6}\pi m \Mp^2} 
\left(\frac{m^2\mathrm{e}^x}
{\lambda v^2 x}\right)^{\frac{\lambda v^2}{2m^2}} 
\Gamma^{\frac{1}{2}}\left(\frac{\lambda v^2}{m^2},
\frac{\lambda v^2}{m^2}x\right)\, .\nonumber\\
\end{eqnarray}
This analytical formula is compared with a numerical integration of the Langevin equations in Fig.~\ref{fig:dispersions}, where the matching appears to be very good in the valley (\ie for $N<N_\uc$). At the critical point where $x=1$, in the limit $\lambda v^2/m^2\gg 1$, one can make use of the asymptotic behavior $(e/y)^y\Gamma(y,y)\simeq\sqrt{\pi/(2y)}$ when $y\rightarrow \infty$ , and the previous expression reduces to
\begin{equation}
\label{eq:sigmapsic}
\sigma_{\chi_\uc}\simeq\left(\frac{\lambda}{2\pi} \right)^{3/4}
\left( \frac{v}{3m}\right)^{1/2}\frac{v^3}{8\Mp^2}\, .
\end{equation}
In the supersymmetric version of the model, where $\Phi_\uc=v$, one then has $\sigma_{\chi_\uc}/v\propto\sqrt{m/\Mp\Delta_\Phi}$, where $\propto$ signals the presence of a numerical $\mathcal{O}(1)$ factor. Since we are working under the $\Delta_\Phi\ll 1$ assumption, for light inflaton fields compared to the Planck mass, this means that $\sigma_{\chi_\uc}\ll v$. Therefore one can safely use the approximation $\chi\ll v$ in Eq.~(\ref{Langphi}), even when the stochastic diffusion in the $\chi$ direction is taken into account. So one can now integrate Eq.~(\ref{Langphi}). If the initial condition for $\varphi$ is chosen to be a delta distribution, this leads to:
\begin{eqnarray}
\varphi&=& \exp{\left[-4\frac{m^2\Mp^2}{\lambda v^4}
\left(N-N_\uin\right)\right]}\left[\varphi_\uin +
2\sqrt{\frac{3}{\lambda}}\frac{\Mp}{v^2}
\right.\nonumber\\& & \quad\quad \left.\times
\int_{N\uin}^{N}\exp\left(4\frac{m^2\Mp^2}
{\lambda v^4}n\right)\xi_\phi\left(n\right)\dd n\right]\, ,
\nonumber\\
\end{eqnarray}
from which one gets a distribution for $\varphi$ centered around its classical counterpart $\left\langle\varphi\right\rangle=\varphi^{(0)}$, with a constant dispersion given by:
\begin{equation}
\label{sigmaphi}
\sigma_\varphi=\frac{\lambda v^4}{8\sqrt{6}\pi m\Mp^2}\, ,
\end{equation}
where Eq.~(\ref{noisecorrpsideSitter}) has been used to obtain the above. In Fig.~\ref{fig:dispersions}, the dispersion given by this formula is compared with the dispersion obtained numerically, from solving $\sim 10^7$ realizations of the Langevin equations. The figure confirms that $\sigma_\varphi$ is indeed constant during the valley phase (\ie for $N<N_\uc$), and shows the good agreement between numerical and analytical methods.

Let us now say a few words about the waterfall phase. The classical dynamics of this phase is reviewed in appendix~\ref{app:waterfall}, where the notations of Ref.~\cite{Kodama:2011vs} are adopted, dividing this stage into three sub-phases, labeled $0$, $1$ and $2$. At the classical level, the sub-phase 2 ends up with the values of the fields given by Eqs.~(\ref{eq:psi2}) and~(\ref{eq:phi2}), which implies that $\chi_2/v\ll 1$ and that $\varphi_2\simeq\Phi_\uc$. Therefore the approximation scheme used for the calculation of the diffusion in the $\chi$-direction in the valley is still roughly valid, as can be confirmed by comparing with the numerical results displayed in Fig.~\ref{fig:dispersions}.

On the other hand, a straightforward way to estimate the dispersion of the distribution for $\chi$ at the end of the sub-phase 2 is to use the following qualitative argument. The stochastic diffusion in the valley phase randomizes which minimum of the potential is eventually taken on by the coarse-grained field, in such a way that half of the Langevin realizations end up in each minimum. If the classical estimation of $\chi_2$ given by Eq.~(\ref{eq:psi2}) is roughly correct, the typical dispersion should be twice this value, namely:
\begin{equation}
\label{sigmapsi2}
\sigma_{\chi_2}\simeq\frac{2m}{g}\, .
\end{equation}
This expression obtained from a heuristic argument is showed to agree with the numerical calculation in Fig.~\ref{fig:dispersions}.

Eq.~(\ref{eq:phi2}) and the fact that $\chi_\uc$ is a stochastic quantity both lead to the conclusion that the value $\varphi_2$ of the inflaton at the end of Phase 1 of the waterfall is also a stochastic quantity. Going back to Eq.~(\ref{eq:phi2}), we see that $\sigma_{\chi_\uc}$ and $\sigma_{\varphi_2}$ can be related to each other. Using the Gaussianity of the distribution for $\chi_\uc$, one obtains:
\begin{eqnarray}
\label{sigmaphi2}
\sigma_{\varphi_2}&=&\Phi_\uc\left\lbrace \exp\left[-\frac{2m}
{\lambda^{1/4}v} \left|\ln\left(\frac{m}
{g\sigma_{\psi_\uc}}\right)\right|^{\frac{1}{2}}\right]
\right.\nonumber\\&-&\left.
\exp\left[-\frac{2m}{\lambda^{1/4}v} \left|\ln\left(\frac{m 
\sqrt{2\pi}}{g\sigma_{\psi_\uc}}\right)\right|^{\frac{1}{2}}\right] 
\right\rbrace^{\frac{1}{2}}\, .
\end{eqnarray}
Again, this value is compared with a numerical calculation in Fig.~\ref{fig:dispersions} which confirms the validity of this approach.

This calculation provides a leading-order result for the field dispersions $\sigma_{\varphi}$ and $\sigma_{\chi}$. To go beyond this approximation, we now proceed to step 3 of our recursive method. 

\section{Linearized Mode Function Calculation}
\label{sec:PertExpAndModeCalc}

We now go back to the mode function equations for the bath fields, and solve them again, this time in the presence of the ``shifted" background calculated in the previous subsection (this now represents keeping corrections up to $\mathcal{O}(\hbar^2)$) and keeping corrections up to leading order in slow-roll.

As shown in Ref.~\cite{paper1}, at this order in $\hbar$, the mode functions we need to solve for correspond to the linear perturbations equations for the scalar fields in a shifted background. We can therefore apply here the usual methods from the theory of linearized cosmological perturbations. This also means that one needs not to worry about mode coupling effects at this order, at least for what concerns the calculation of the bath propagators and noise amplitudes\footnote{However, technically, at this order in $\hbar$, we should include the loop corrections to the Langevin equations calculated in Ref.~\cite{paper1}, which would come in at the fourth and last stage of the recursive method applied in the present paper. Despite this, since these effects represent mode coupling between bath and system fields, we expect them to be negligible in the valley phase of inflation. We therefore neglect them at this order and plan on coming back to this calculation in a future focused on the waterfall phase of inflation, where those effects are known to be important (see e.g. Ref.~\cite{Levasseur:2010rk})}.

From the split of the full fields into bath and system in (\ref{phibathsystsplit}) and (\ref{psibathsystsplit}), one can think about this as performing the following expansion in the bath fields:
\begin{eqnarray}
\phi_>\left(x,N\right)&=&\delta\phi_>^{\left(1\right)}\left(x,N\right)
+\delta\phi_>^{\left(2\right)}\left(x,N\right)+\cdots
\, ,\nonumber \\
\psi_>\left(x,N\right)&=&\delta\psi_>^{\left(1\right)}\left(x,N\right)
+\delta\psi_>^{\left(2\right)}\left(x,N\right)+\cdots\, ,\nonumber\\
\end{eqnarray}
where the zero mode of the bath fields is zero, by definition. The background quantities are given by the homogeneous coarse-grained fields $\varphi^{(0)}$ and $\chi^{(0)}$, and we aim at solving for the first order fluctuations, $\delta\phi_>^{\left(1\right)}$ and $\delta\psi_>^{\left(1\right)}$. The different notations refering to the different quantities derived from the quantum fields $\Phi$ and $\Psi$ are summed up in appendix \ref{app:NotationsAndAssumptions}. Following the recursive scheme presented above, in the equations of motion driving these quantities, all the occurrences of the coarse grained quantities $\varphi^{(0)}$ and $\chi^{(0)}$, which are stochastic quantities, shall be replaced by their stochastic means, namely
\begin{equation}
F\left[\varphi^{(0)},\chi^{(0)}\right]\rightarrow\left\langle 
F\left[\varphi^{(0)},\chi^{(0)}\right] \right\rangle\, ,
\end{equation}
where $F$ is any functional of the two fields. Moreover, in order for this expansion to be consistent, we also need to include metric perturbations. In the following, we will only be interested in the scalar mode and will therefore neglect the tensor modes. We choose to work in the uniform curvature gauge at linear order in metric perturbations.

\subsection{First-Order Metric Perturbations}

For the Friedman - Lema\^ itre - Robertson - Walker (FLRW) metric at linear order, scalar, vector and tensor metric perturbations decouple (see Refs~\cite{Mukhanov:1990me, Brandenberger:2003vk, Malik:2008im, Baumann:2009ds} and references therein for a review of the theory of cosmological perturbations). We therefore need only to consider scalar perturbations at this order. For a  flat FLRW space-time, they are parametrized by:
\bea
	\dd s^2&=& -(1+2\alpha)\dd t^2-aB_{,i}\dd t \dd x^i\nonumber\\
	& & +a^2[\delta_{ij}(1-2\beta)+E_{,ij}]\dd x^i \dd x^j\, .
\eea
Using the conventions of Ref.~\cite{Finelli:2003bp}, in the following we work in the spatially-flat, or uniform curvature, gauge, which is defined by making the scale factor of the metric homogeneous choosing $\beta=E=0$:
\bea
	\dd s^2= -(1+2\alpha)\dd t^2-aB_{,i}\dd t\dd x^i+a^2 \dd x^2\, .
\eea
This choice uniquely fixes the gauge. The Einstein equations then reduce to:
\bea
\label{generaleinsteineqnsinflatgaugeIn}
	3H^2\alpha+\frac{k^2}{2a^2}(aHB)&=& 
	-\frac{\delta \rho}{2\Mp^2}\, ,\\
	H\alpha&=&-4\pi G\delta q\, ,\\
	H\dot \alpha +(3H^2+2\dot H)\alpha&=& \frac{1}{2\Mp^2}
	\left( \delta p - \frac{2}{3}k^2\delta \Sigma \right)\, ,\\
	(\partial_t+3H)\frac{B}{2a}-\frac{\alpha}{a^2}&=&
	\frac{\delta \Sigma}{2\Mp^2}\, ,
\label{generaleinsteineqnsinflatgaugeEnd}
\eea
where $\Sigma$ stands for the anisotropic stress, which we set to zero from now on since it cannot be seeded by scalar field matter to linear order in perturbation theory, and the total density and momentum perturbations are given by:
\bea
	\delta \rho &=& \dot{\varphi^{(0)}}\left( \delta\dot{\phi}^{(1)}_>
	-\dot\varphi^{(0)}\alpha \right)+\dot{\chi}^{(0)}
	\left(\delta\dot{\psi}^{(1)}_>-\dot\chi^{(0)}\alpha \right)
	\nonumber\\
	&&+V_{,\Phi}(\varphi, \chi)\delta\phi^{(1)}_>+V_{,\Psi}
	(\varphi, \chi)\delta\psi^{(1)}_>\, ,\\
	\delta q &=&
    \dot\varphi^{(0)}\delta\phi^{(1)}_>
    +\dot\chi^{(0)}\delta\psi_>^{(1)}\, ,
\eea
where $V_{,\Phi}$ and $V_{,\Psi}$ stand for the derivatives of the potential with respect to the fields $\Phi$ and $\Psi$, evaluated at their coarse-grained values. In order to obtain equations for $\phi_>$ and $\psi_>$ only, one just needs to consider the first two of the Einstein's equations in Eqs.~(\ref{generaleinsteineqnsinflatgaugeIn}-\ref{generaleinsteineqnsinflatgaugeEnd}), that is the $G_0^0$ and the $G_i^0$ equations, which can be expressed as:
\bea
	-\frac{H}{a}k^2B&=&8\pi G\left[\dot \varphi^{(0)} \delta\dot 
	\phi^{(1)}_>+V_{,\Phi}\delta\phi_>^{(1)}\right.\nonumber\\
	&&\qquad\left.+\dot \chi^{(0)} \delta\dot \psi^{(1)}_>
	+V_{,\Phi}\delta\psi_>^{(1)}+2V\alpha \right]\nonumber\\
	&=&\frac{8\pi G}{H}\left\lbrace\dot\varphi^{(0)^2}\frac{\dd}
	{\dd t}\left[\frac{H\delta\phi_>^{(1)}}
	{\dot \varphi^{(0)}}\right]\right.\nonumber \\
	\label{constraintequationsGRflucB}
	&& \qquad\qquad\left.+\dot\chi^{(0)^2}\frac{\dd}
	{\dd t}\left[\frac{H\delta\psi_>^{(1)}}
	{\dot\chi^{(0)}}\right]\right\rbrace\, ,\\
	\alpha_{,i}&=& \frac{4\pi G}{H}\left[ 
	\dot \varphi^{(0)}\delta\phi^{(1)}_{>_{,i}} 
	+\dot \chi^{(0)}\delta \psi_{>_{,i}}^{(1)}\right]\, ,
\label{constraintequationsGRflucAlpha}
\eea
where $G=1/(8\pi\Mp^2)$ is the gravitational constant. Also, since we are assuming the absence of anisotropic stress, we have the extra constraint $\dot B+2HB=2\alpha/a$ (which is the equivalent of the usual $\Phi=\Psi$ equality in the longitudinal gauge). 

\subsection{Inflaton Fluctuations $\delta\phi^{(1)}$}
\label{sec:deltaphi}

Following the recursive strategy presented above, let us now write down \cite{Gordon:2000hv} the equation of motion for the first order inflaton fluctuations $\delta\phi^{(1)}_>$, replacing the functions of the background fields $\varphi^{(0)}$ and $\chi^{(0)}$ by the stochastic mean values of the same functions of the coarse grained quantities $\varphi$ and $\chi$:
\bea
	\delta 
	\ddot{\phi}_{\bm{k}}^{(1)}+3H\delta\dot\phi_{\bm{k}}^{(1)}
	+\left(\frac{k^2}{a^2}+m^2+g^2\left\langle\chi^{2}\right\rangle  
	\right)\delta\phi_{\bm{k}}^{(1)}\quad\nonumber  \\ 
	+2g^2\left\langle\varphi\chi\right\rangle 
	\delta\psi_{\bm{k}}^{(1)}\qquad\qquad \qquad\qquad\qquad 
	\qquad\nonumber \\
	= 2\alpha \left\langle\ddot{\varphi}\right\rangle
	+\left\langle\dot\varphi\right\rangle\left(\dot \alpha 
	+6H\alpha+\frac{k^2}{2a}B\right)\, .
\label{phi1}
\eea
The notation ``${}_>$'' has been dropped for notational simplicity. As derived above, the distribution for $\varphi$ is centered around its classical counterpart $\left\langle\varphi\right\rangle=\varphi^{(0)}$ and therefore, one can replace $\left\langle\dot{\varphi}\right\rangle = \dot{\varphi}^{0}$ and $\left\langle\ddot{\varphi}\right\rangle = \ddot{\varphi}^{0}$. Then, for $\Delta_\Phi\ll 1$, assumption under which we are currently working, the noise effects in the $\varphi$ direction do not affect much the inflaton dynamics, {\it i.e.} $\sigma_\varphi/\varphi\ll 1$. Assuming independence of the two coarse-grained field probability density functions, one can then approximate $\left\langle\varphi\chi\right\rangle \simeq \left\langle\varphi\right\rangle\left\langle\chi\right\rangle = \varphi^{(0)}\left\langle\chi\right\rangle = 0$, the last approximation justified by the fact that the $\chi$ distribution is quickly centered around $0$ in the valley phase\footnote{This approximation is no longer valid in the waterfall phase.}. Replacing finally $\left\langle\chi^{2}\right\rangle$ by $\sigma_\chi^2$, one obtains 
\bea
	\delta\ddot{\phi}_{\bm{k}}^{(1)}+3H\delta\dot\phi_{\bm{k}}^{(1)}
	+\left(\frac{k^2}{a^2}+m^2+g^2\sigma_\chi^2 \right)
	\delta\phi_{\bm{k}}^{(1)}\nonumber  \\
	= 2\alpha \ddot{\varphi}^ {(0)} + \dot\varphi^{(0)}\left(\dot 
	\alpha +6H\alpha+\frac{k^2}{2a}B\right)\, .
\eea

One can see that, in general, the inflaton and the waterfall fields also couple through the metric perturbations on the right hand side. Indeed, since there really are only two degrees of freedom in the problem, it is possible to replace the metric fluctuations in favor of the fields using the constraint equations (\ref{constraintequationsGRflucB}-\ref{constraintequationsGRflucAlpha}). In this process, we set terms with odd powers of $\chi^{(0)}$ to zero, while terms with a quadratic power of $\chi^{(0)}$ to $\langle \chi^{2}\rangle=\sigma_\chi^{2}$. We obtain:
\bea
	\delta\ddot{\phi}_{\bm{k}}^{(1)}+3H\delta\dot\phi_{\bm{k}}^{(1)}
		+\qquad \qquad \qquad\qquad \qquad \qquad \\
		\left[\frac{k^2}{a^2}+m^2+g^2\sigma_\chi^2-\frac{8\pi G}
		{a^3}\frac{\dd}{\dd t}\left(\frac{a^3\dot{\varphi}^
		{(0)^2}}{H} \right)  \right]\delta\phi_{\bm{k}}^{(1)}=0.
		\nonumber  \\
\eea
Here, the last term is clearly identifiable as coming from gravitational interactions since it is proportional to the gravitational constant. One also sees that, written in this way, the waterfall field seems to decouple from the inflaton field. Indeed, this same equation would have been obtained for a single scalar field (with a stochastically-shifted mass) coupled to the metric perturbations. 

This equation can also be rewritten in a way that makes explicit of what order in slow-roll the gravitational corrections are, and let the corrections coming from $\sigma_\chi$ appear clearly:
 \bea
		\delta \ddot{\phi}_{\bm{k}}^{(1)}+3H\delta\dot\phi_{\bm{k}}^{(1)}+	\Bigg[\frac{k^2}{a^2}+m^2+g^2\sigma_\chi^2+2\frac{\dot H}{H}\times\nonumber   \\
	\left( \frac{\ddot{\varphi}^{(0)}}{\dot{\varphi}^{(0)}}-\frac{\dot H}{H}+3H  \right)\left(1+ \frac{1} {\dot \varphi^{(0)^2}/\dot \sigma_\chi^{2}+1}\right)\Bigg]\delta\phi_{\bm{k}}^{(1)}=0\, .\nonumber\\
	\label{phiwithmetricpertsBE}
\eea
From Eq.~(\ref{sigmapsi}), one can calculate the time variation of $\sigma_\chi$ at the critical point 
\be 
\left.\dd \sigma_\chi/\dd N\right\vert_{\uc}=\left(2\pi\right)^{5/4}\lambda^{1/4}\sqrt{mv/3}\, .
\ee 
From this one obtains a typical value
\be 
\dot \varphi^{(0)^2}/\dot \sigma_\chi^{2} \simeq 192 \sqrt{2}\pi^{5/2}m^3\Mp^4/ \left(g^2\lambda^{3/2}v^7\right)
\ee
which is typically very big (\eg for the parameters values used in Fig.~\ref{fig:dispersions}, one obtains $\simeq 0.5\times 10^{6}$). One can therefore approximate the second parenthesis term of the previous equation to be $\simeq 1$.

In Eq.~(\ref{phiwithmetricpertsBE}) also, the $H$ factors should be understood as $\left\langle H\left(\varphi,\chi\right)\right\rangle_{\xi_\varphi,\xi_\chi}$ and similarly for any function of $H$ ($\dot H/H$, \textit{etc}), and more generally any function of coarse grained quantities. However, in the valley $H$ is assumed to be vacuum dominated, and its time-dependence mostly comes from $\varphi\simeq\varphi^ {(0)}$. The Hubble parameter can therefore be treated in the standard way without impacting much on the result. This is why a lighter notation is adopted for this parameter.

Since $\left(\dot \varphi^{(0)}\right)^2$ dominates the contribution to $\dot H$, one finds that the corrections due to $\chi$ are negligible and recovers that the metric perturbations cause a shift in the mass of a single field coupled to the metric. The effective mass for the inflaton can therefore be rewritten in terms of the first slow-roll parameter:
\bea
	m^2+g^2\sigma_\chi^2 +2\frac{\dot H}{H}\left( \frac{\ddot{\varphi}^{(0)}}{\dot{\varphi}^{(0)}}-\frac{\dot H}{H}+3H  \right)\qquad\qquad\qquad\\
	\approx m^2+g^2\sigma_\chi^2 -6H^2\left( \varepsilon_1-\frac{1}{3}\varepsilon_1^2 +\frac{\dot \varepsilon_1}{3H}  \right)\, .\nonumber
\eea

Upon the standard field redefinition to obtain the canonically normalized field
\be 
\delta\phi^{(1)}_{\bm{k}}=a^{-1}v_{\bm{k}}
\ee
and the change of the time coordinate to conformal time $\dd\tau=a^{-1}\dd t$, one obtains an equation analogous to the usual mode function for a single scalar field in de-Sitter space:
\bea
	v''_{\bm{k}}+\left\lbrace k^2-\frac{a''}{a}\right.\qquad\qquad\qquad\qquad\qquad\qquad\qquad \\
	\left.+a^2\left[m^2+g^2 \sigma_\chi^{2} -6H^2\left( \varepsilon_1-\frac{1}{3}\varepsilon_1^2 +\frac{\dot \varepsilon_1}{3H}  \right)\right] \right\rbrace v_{\bm{k}}=0\, ,\nonumber
\eea
where the prime denotes a derivative with respect to the conformal time $\tau$. Or, to first order in slow-roll (which we assume is sufficient in the valley), $aH=-\tau^{-1}(1-\varepsilon_1)$, and
\be
\label{modefunction}
	v''_{\bm{k}}+\left[ k^2-\frac{2-m^2/H^2-g^2 \sigma_\chi^{2}/H^2+9\varepsilon_1}{\tau^2} \right]v_{\bm{k}}=0\, .
\ee

We can then quantize the modes by promoting $v_{\bm{k}}$ to an operator 
\be 
\hat{v}_{\bm{k}}(\tau)=v_{\bm{k}}(\tau)\hat{a}_{\bm{k}} +v_{\bm{k}}^{\star}(\tau)\hat{a}^\dagger_{-\bm{k}}
\ee
and imposing the usual commutation relations 
\be 
\left[ a_{\bm{k}}, a^\dagger_{-\bm{k}'}\right]=(2\pi)^3\delta^{(3)}(k+k') \, .
\ee
Noticing that, from Eq.~(\ref{sigmapsi}), one has 
\be 
\left.\dd/\dd N \left(\sigma_\chi^2 / H^2\right) \right\vert_\uc= 1 /(4\pi^2)
\ee
at the critical point, the time variation of the $\sigma_\chi^ 2$ in the above equation is suppressed by a $g^2$ factor and can be neglected in the adiabatic limit, allowing us to express the solution to the mode function in terms of Hankel functions:
\be
\label{eq:vk}
	v_{\bm{k}}=-ie^{i(\nu+\frac{1}{2})\frac{\pi}{2}}\frac{\sqrt{\pi}}{2}(-\tau)^{1/2}H_{\nu}^{(1)}(-k\tau),
\ee
where 
\bea 
\nu^2 &=&9/4-(m^2+g^2\sigma_\chi^{2})/H^2+9\varepsilon_1 \nonumber \\
&\approx& \frac{9}{4}+\frac{3}{2}\varepsilon_2+\frac{g^2\sigma_\chi^{2}}{H^2}+9\varepsilon_1 \, .
\eea
In the second line, we have re-introduced the second slow-roll parameter to make explicit which corrections in slow-roll we are keeping. This term is the term which propagates to yield the well-known classical blue tilt for the canonical hybrid inflation model. The last term is the correction from metric fluctuations which induces a red tilt. The second-to-last term, however, is a new term which is induced by stochastic effects and which tends to increase the blue tilt.

The mode functions have been normalized so that deep inside the Hubble radius, when the $k^2$-term dominates the mass in Eq.~(\ref{modefunction}), one recovers the Bunch-Davies vacuum: 
\be
	v_{\bm{k}}\rightarrow\frac{e^{-ik\tau}}{\sqrt{2k}}, \qquad \tau\rightarrow-\infty\, .
\ee
Here a few comments are in order. First, note that what we have calculated so far are only the linearized mode functions of the bath quantum fields, not the perturbations that will arise in the coarse-grained system once we perturb the Langevin equations, and which are the ones giving rise to the classical curvature perturbations. As so, to be technically correct we are not allowed yet to predict the modified spectral index (even though we can suspect that the result we obtain here should propagate to the final answer). We first have to use this corrected amplitude of linearized mode functions to calculate a shifted noise through Eq.~(\ref{noisecorrphi}), and then use the latter to source a new solution to the Langevin equation (\ref{Langphi}). Linear perturbations around this classical system will allow us to predict $n_s$ to leading order in slow-roll.

Second, note that the effect of $\sigma_\chi^{2}$ on $\delta \phi_{\bm{k}}^{(1)} $ is to make each mode more massive. Therefore, having the sub-Hubble modes evolve in a background that has been shifted by the integration to first order of all modes which have already frozen out has the effect of making the tilt of the inflaton modes bluer when they freeze out. It should be highlighted that this conclusion does not depend on the specific value of $\sigma_\chi$, and will remain true when its calculation is refined in section~\ref{sec:sigmapsi}. Moreover, for typical values of the potential (and in particular in the supersymmetric version of the model $\Phi_\uc \sim v$), one has 
\be 
g^2 \sigma_\chi^{2}/H^2 \sim \varepsilon_1^{-1/4}
\frac{\lambda v^3}{\Phi_{\mathrm{c}}^{3/2} \Mp^{3/2}} \gg 9\varepsilon_1 \, .
\ee
Therefore, the blue tilt induced by the stochastic background will always overcome the tendency of metric perturbations to make the spectrum of quantum fluctuations red. 

This result is not \apriori obvious since the two effects are antagonist (the coupling to metric perturbations rendering the spectral tilt redder and the stochastic shift of the background rendering it bluer). It is necessary to rigorously work out the two contributions in order to conclude that the latter wins over the former,  yielding a shifted and a blue-tilted spectrum of the quantum noise sourcing the Langevin equations once the mode functions are plugged back in equations (\ref{noisecorrphi}) and (\ref{noisecorrpsi}).

We once again insist that whether this blue shift and time dependence in the noise amplitude also yields a worsened blue tilt problem, by translating into a bluer spectrum of classical curvature perturbations of the coarse-grained field $\varphi$ (which are the observable ones), is a different question which requires further calculation. To provide a satisfactory answer, we shall wait until we feed this new quantum noise amplitude back into the Langevin equations (\ref{Langphi}) and (\ref{Langpsi}) and calculate the spectrum.

As a second remark, note that if the collective effect of the inflaton mass and $\sigma_\chi$ is a small enough correction, {\it i.e.} if $(m^2+g\sigma_\chi^2)/H^2-9\varepsilon_1\leq 9/4$, then as the modes $v_{\bm{k}}$ cross their Hubble radius, their oscillations stop and they freeze out as one would expect. However the modes of the original field $\delta\phi_{\bm{k}}^{(1)}$ also contain a decay factor $a^{-1}(-k\tau)^{-\nu+1/2}\sim \tau(-k\tau)^{-\nu+1/2}$ as $-k\tau\rightarrow 0$, which indicates that they eventually roll back down to zero. (Recall that the observable quantity here is the curvature perturbation, $\mathcal{R} = \frac{H}{\dot{\phi}}\delta\phi$). This means that the modes become over-damped after horizon exit. The full solution in this limit is given by:
\bea
	v_{\bm{k}}\rightarrow\begin{cases}-e^{i(\nu+\frac{1}{2})\frac{\pi}{2}}\frac{2^{\nu-1}}{\sqrt{\pi}}\Gamma(\nu)\frac{(-\tau)^{-\nu+1/2}}{k^{\nu}} \qquad 0<\nu\leq3/2\\ e^{i\frac{\pi}{2}}(-\tau)^{1/2}\ln(-k\tau)\qquad \qquad \qquad \qquad \nu=0\end{cases}
\eea
The $k^\nu$ factor shows the deviation from scale invariance, and we therefore recover that the mass of the inflaton causes the spectrum to be blue-tilted in the valley (scale invariance has $k^{3/2}$, which is the massless case). The power of $\tau$ shows the time dependence, and in the massless case one recovers $\tau^{-1}$, which is canceled by multiplying by $a^{-1}$ to recover $\delta\phi_{\bm{k}}^{(1)}$.

\subsection{Waterfall Fluctuations $\delta\psi^{(1)}$}
\label{sec:deltapsi}

\subsubsection{Mode function evolution equation}

Let us now proceed with $\delta\psi^{(1)}$, similarly expanding the equations of motion to first order, once again in the flat-slicing gauge. One obtains:
\bea	\delta\ddot{\psi}_{\bm{k}}^{(1)}+3H\delta\dot{\psi}_{\bm{k}}^{(1)}+\left(\frac{k^2}{a^2}+3\lambda\chi^{2}-\lambda v^2 +g^2\varphi^{2}\right)\delta\psi^{(1)}_{\bm{k}}\nonumber\\
	+2g^2\varphi\chi\delta \phi^{(1)}_{\bm{k}}=\dot \alpha\dot \chi-2\alpha V_{,\Psi}\left(\varphi,\chi\right)-\dot{\varphi}\frac{k^2}{2a}B\, .\quad\qquad
\label{psi1}
\eea
As in the previous subsection, on the left hand side, one replaces $\chi^2$ by $\sigma_\chi^2$, $\varphi^2$ by $\left\langle\varphi^2\right\rangle\simeq{\varphi^{(0)}}^2$, and $\varphi\chi$ by $\left\langle\varphi\chi\right\rangle=0$. On the right hand side, using the linearized Einstein equations to replace the metric fluctuations by field perturbations, and setting to zero all terms with stochastic mean values with an odd powers of $\chi$ (remembering that the distribution of $\chi$ is even), one obtains:
\bea
	\delta\ddot{\psi}_{\bm{k}}^{(1)}+3H\delta\dot{\psi}_{\bm{k}}^{(1)}+\left[\frac{k^2}{a^2}+3\lambda \sigma_\chi^{2}-\lambda v^2+g^2\varphi^{(0)^2} \right.\nonumber
	\\\left.-\frac{8\pi G}{a^3}\frac{\dd}{\dd t}\left(\frac{a^3\left\langle\dot{\chi}^{2}\right\rangle}{H} \right) \right]\delta\psi^{(1)}_{\bm{k}}=0\, ,\qquad
\eea
where again, $H$ is approximated by its classical value $H(\varphi^{(0)},\chi=0)$. Note that, as opposed to what would have been obtained using perturbations theory around a classical background for $\Psi$, the stochastically shifted background causes the $\delta\psi^{(1)}_{\bm{k}}$ perturbations not to decouple completely. Indeed, in the case of a classical background, unless the trajectory is turning in field space, the perturbations reduce to those of a scalar field in an unperturbed FLRW space-time \cite{Gordon:2000hv}. This is not the case here: the field space trajectory is straight, but the stochastic dispersion of the waterfall allows for non-vanishing corrections due to gravity.

We again rewrite the term coming from gravitational interactions in terms of the slow-roll parameters:
\bea
	&2\frac{\dot H}{H}&\left( \frac{\ddot{\chi}}{\dot{\chi}}-\frac{\dot H}{H}+3H  \right)\left(1+ \frac{1}{\dot \sigma_\chi^{2}/\dot \varphi^{(0)^2}+1}\right)\simeq \nonumber \\
	& & 4\frac{\dot H}{H}\left(-\frac{\dot H}{H}+3H \right)
	= 12H^2\left(\varepsilon_1- \frac{1}{3}\varepsilon_1^2\right)\, .
\eea
As above, one proceeds to the field redefinition 
\be 
\delta\psi^{(1)}_{\bm{k}}=a^{-1}u_{\bm{k}}
\ee 
and changes coordinates to conformal time, $\dd t=a\dd \tau$, to find the mode function expressed in terms of the canonical variable $u_{\bm{k}}$: 
\bea
	u_{\bm{k}}''+\left\lbrace k^2-\frac{a''}{a}+a^2\left[3\lambda\sigma_\chi^2-\lambda v^2 +g^2\varphi^{(0)^2}\right.\right.\\
	 \left.\left.+12H^2\left(\varepsilon- \frac{1}{3}\varepsilon^2\right) \right]\right\rbrace u_{\bm{k}}=0\, .\nonumber
\eea
Using the explicit expression for $a$ during inflation to first order in slow-roll, $aH=-\tau^{-1}(1-\varepsilon_1)$, and under the assumption of vacuum domination, this gives rise to
\bea
\label{ukgenericmodefunction}
	u_{\bm{k}}''+\Bigg[\left. k^2-\frac{1}{\tau^2}\times\right. \qquad\qquad \qquad \qquad\qquad \qquad \qquad \qquad\\
	\left. \left(2-3\frac{\lambda\sigma_\chi^2}{H^2} + \frac{12\Mp^2}{v^2}  -\frac{g^2\varphi^{(0)^2}}{H^2} +15\varepsilon_1\right)\right. \Bigg]u_{\bm{k}}=0\, . \nonumber
\eea
In contrast to what happens for the fluctuations of the rescaled inflaton field, the correction terms in the mode equation for the fluctuations of the waterfall field are large. This is a reflection of the tachyonic instability in the direction of the waterfall field. More specifically, the mass term of this equation contains terms of different orders of magnitude. Indeed, in the vacuum dominated regime, under the slow-roll approximation, and since $\sigma_\chi^2<\sigma_{\chi_\uc}^2$ in the valley, one has
\be 
 15\varepsilon_1\, , \, 3\frac{\lambda\sigma_\chi^2}{H^2} \ll 2 \ll \frac{12\Mp^2}{v^2} \, , \, \frac{g^2\phi^{(0)^2}}{H^2} \, .
\ee
Moreover, for typical parameter values, one also has:
\be
	15\varepsilon_1\, \ll \, 3\frac{\lambda\sigma_\chi^2}{H^2}
	\, ,
\ee
although it would in principle be possible to find a range of fine tuned parameters for which this inequality does not hold.

Finally, one can use Eq.~(\ref{Nc}) to rewrite $\varphi_{\mathrm{in}}^{(0)}$ in terms of $\Phi_\uc$ and the total number of $e$-folds of inflation $N_\uc$ produced in the valley when $\varphi$ crosses the critical point, and one obtains
\bea
\label{modefctmassPsi1-phidep}
	\frac{g^2\varphi^{(0)^2}}{H^2}&=&\frac{12\Mp^2}{v^2}\frac{\varphi^{(0)^2}}{\Phi_{\uc}^2}\\
\label{modefctmassPsi2-phidep}
	&\simeq & \frac{12\Mp^2}{v^2}\left[e^{N_c-N(\tau)}\right]^{8\Mp^2m^2/(v^4\lambda)},
\eea
where at first order in $\varepsilon_1$ one has 
\be 
N(\tau)=\ln\left[-(1+\varepsilon_1)/(H\tau)\right]\approx-\ln(-H\tau)+\varepsilon_1 \, ,
\ee
if one initializes $N$ to $0$ when  $\tau=-(1+\varepsilon_1)/H$.

\subsubsection{Qualitative mode evolution analysis}

Let us now try to gain some qualitative insight about the time-evolution of the $\bm{k}$-modes. First of all, one can see that there is some explicit $\tau$-dependence in the time-dependent mass of the $u_{\bm{k}}$'s (through the $\phi^{(0)^2}$) in addition to the usual $1/\tau^2$ dependence. One needs to make sure that this term goes to zero at early times, \ie when the limit $k\tau\rightarrow -\infty$ is formally taken, so that the Bunch-Davies vacuum initial condition can be recovered in that limit. That is, one needs to make sure that at arbitrary early times (as $k\tau\rightarrow-\infty$), any given mode is at small enough scales so that it feels a Minkowski flat space-time and lies in the Bunch-Davies state.

Bearing this in mind, since $8\Mp^2m^2/(v^4\lambda)\ll 1$, from our assumption of vacuum-domination of the Hubble constant in the valley, one is safe since
\be
	\frac{(-H\tau)^{\frac{8\Mp^2m^2}{v^4\lambda}}}{\tau^2}\rightarrow 0 \qquad \mathrm{as} \qquad (-k\tau)\rightarrow \infty\quad .
\ee
One can therefore quantize the mode functions as usual using the Bunch-Davies vacuum solution as a limiting initial condition at early times. 

Also, the $\varphi^{(0)}$-dependence of the mass was written in the form (\ref{modefctmassPsi1-phidep}) in order to get a better insight on the qualitative behavior of the modes after they exit the Hubble radius. Inserting this expression into the mode functions equations of motion, one obtains:
\be 
\label{effectivemass0}
	u_{\bm{k}}''+\left[k^2-m_u^2\left(\tau\right)\right]u_{\bm{k}}=0\, ,\\
\ee
where the effective mass $m_u$ is defined as
\bea
\label{effectivemass1}
	m_u^2(\tau)&\equiv&\frac{2-m_\psi^2/H^2}{\tau^2}\\&=&
	\frac{1}{\tau^2}\left[2+15\varepsilon_1- 3\frac{\lambda\sigma_\chi^2}{H^2} - \frac{12\Mp^2}{v^2} \left( \frac{\varphi^{(0)^2}}{\varphi_{\uc}^2}-1\right) \right]\, .\nonumber\\
\eea

\begin{figure}
\begin{center}
\includegraphics[width=9cm]{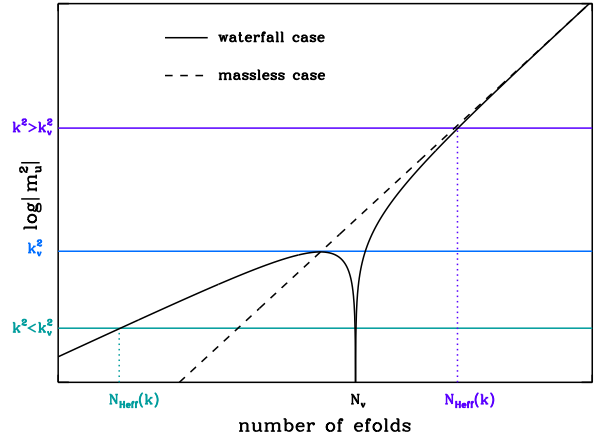}
\caption{Sketch of the time evolution of the squared mass $m_u^2$ appearing in Eq.~(\ref{effectivemass0}), as a function of the number of $e$-folds. As a comparison, the dashed line represents the massless situation where $m_\psi=0$ and $m_u^2=2/\tau^2$. For $N<N_{\mathrm{v}}$ (respectively $N>N_{\mathrm{v}}$), one has $m_u^2<0$ (respectively $m_u^2>0$). The first time a mode $\bm{k}$ crosses the squared mass scale is given by $N_{H_{\mathrm{eff}}}\left(k\right)$. Its behavior then depends on whether $N_{H_{\mathrm{eff}}}\left(k\right)<N_{\mathrm{v}}$ or $N_{H_{\mathrm{eff}}}\left(k\right)>N_{\mathrm{v}}$ (see discussion in the text). The separation between these two regimes is given by $k_\mathrm{v}$.}
\label{fig:psimasssketch}
\end{center}
\end{figure}

The time evolution of the squared mass $m_u^2$ is sketched in Fig.~\ref{fig:psimasssketch}. Very small scales for which $k^2\gg m_u^2(\tau)$ are still oscillating in their Bunch-Davies state. However when a mode crosses the value $k_{H_\mathrm{eff}}^2=m_u^2(\tau)$, its qualitative behavior changes. We call this point the crossing of the ``effective'' Hubble radius (since in standard massless-single-field inflation this corresponds to the point where every mode crosses the Hubble radius and then freezes out). The evolution of modes with wavelength larger than this effective Hubble crossing scale, \ie satisfying $k^2 < k_{H_\mathrm{eff}}^2 = m^2_u(\tau)$, will be driven according to the behavior of $m^2_u(\tau)$.

Let us see in more detail how the evolution proceeds. The time of ``effective'' Hubble radius crossing $N_{H_\mathrm{eff}}\left(k\right)$ is defined by $k^2=m_u^2\left[N_{H_\mathrm{eff}}\left(k\right)\right]$. This happens during the valley phase if $N_{H_\mathrm{eff}}<N_{\mathrm{c}}$, \ie for modes such that $k<k_\mathrm{c}$, where
\be 
  k_\mathrm{c}^2=m_u^2\left(\tau_\uc\right)\simeq 2H^2\mathrm{e}^{2N_\uc}\, .
\ee
The comoving wave-number $k_c$ thus corresponds to the wavelength that freezes out when $\varphi=\Phi_{\uc}$. Now, for $k<k_\mathrm{c}$, two different behaviors for $m_u^2$ right after effective Hubble radius crossing can occur, which we now investigate. Recall that typically one has $\frac{12\Mp^2}{v^2}\gg 2 \gg 3\frac{\lambda\sigma_\chi^{2}}{H^2}$. Therefore the modes such that $m_u^2<0$, when they cross their effective Hubble radius, are such that 
\bea
N_{\uc}-N_{H_\mathrm{eff}}\left(k\right) 
&>& \frac{v^4\lambda}{8\Mp^2m^2}\ln(1+\frac{v^2}{6\Mp^2}) \nonumber \\
&=& \frac{v^2H^2}{4\Mp^2m^2}+\mathcal{O}(\frac{v^4}{\Mp^4}) \, . \nonumber
\eea
This means that it happens at long wavelength, for $k<k_\mathrm{v}$, where
\be
k_\mathrm{v}^2=m_u^2\left(\tau_{\mathrm{v}}\right)
\simeq \frac{H^2 v^2}{6\Mp^2}
\mathrm{e}^{2 N_\uc-\frac{v^2H^2}{2\Mp^2m^2}}\, .
\ee
From this formula it is very easy to check that $k_{\mathrm{v}}<k_{\uc}$, since $v/\Mp\ll 1$ for this model to proceed at small fields. If $k<k_\mathrm{v}$, the modes do not freeze out as they escape their effective Hubble radius, but rather continue their oscillations until the time $N_\mathrm{v}$ given by 
\be 
\label{eq:Nv}
N_\mathrm{c}-N_\mathrm{v}=\frac{v^2H^2}{4\Mp^2m^2}
\ee
when the distance between $\varphi^{2}$ and its critical value becomes such that $\varphi^{2}- \Phi_{\uc}^2\lesssim ( v^2 /6\Mp^2)\Phi_{\uc}^2$, at which point they freeze out. Mapping back to $\delta\psi^{(1)}_{\bm{k}}=a^{-1}u_{\bm{k}}$, this oscillation period for the $u_{\bm{k}}$ modes  corresponds to a fast under-damping of the perturbations $\delta\psi^{(1)}_{\bm{k}}$. This means that those modes do not undergo squeezing and therefore do not experience classicalization. 

From Fig.~\ref{fig:psimasssketch}, one can see that between $N_{H_\mathrm{eff}}$ and $N_\mathrm{v}$, these modes actually experience a very brief stage during which $k^2$ dominates over $m_u^2$ again. Since this period is very short in time, we will neglect its effect for now (and this approximation will be shown to be consistent \aposteriori). Finally, in the $N\ll N_{\mathrm{v}}$ limit, one has $m_u^2 \simeq -12/\tau^2 \Mp^2/v^2 {\varphi^{(0)}}^2 / \Phi_\uc^2$, which leads to
\bea
N_{H_{\mathrm{eff}}}&\simeq & 
\frac{\log\left(\frac{v^2}{\Mp^2}\frac{k^2}{12H^2}\right)}
{2\left(1-4\frac{m^2\Mp^2}{\lambda v^4}\right)}
-\frac{4}{\frac{\lambda v^4}{2m^2\Mp^2}-2}N_\uc
\cr
&\simeq& \frac{1}{2}\log\left(\frac{v^2}{\Mp^2}\frac{k^2}{12H^2}\right)
-8\frac{m^2\Mp^2}{\lambda v^4}N_\uc\, ,
\eea
the second expression being simplified using the slow-roll condition $\varepsilon_1\left(\Phi_\uc,\Psi=0\right)\ll 1$. 

On the other hand, if $k_\mathrm{v}<k<k_\mathrm{c}$, the modes freeze out and cease to oscillate right after effective Hubble crossing. They consequently undergo squeezing, which allows for classicalization. This phase typically extends over much more than 60 $e$-folds before the inflaton reaches its critical value, which makes it the most important one to study. Close to the inflaton critical value,  $m^2_u(\tau)$ is dominated by the $2/\tau^2$ term, and therefore modes freezing out and growing in this range of conformal time behave very similarly to the perturbations of a light scalar field in de-Sitter space, with a slight positive mass given by $3\lambda\sigma_\chi^2 / H^2-15 \varepsilon+ 12\Mp^2/ v^2\left({\varphi^{(0)}}^2/ {\Phi_\uc}^2 -1\right)$, which gives the spectrum a blue tilt. In this limit where $N_{\mathrm{v}}\ll N<N_\mathrm{c}$, one has
\bea 
N_{H_\mathrm{eff}}\left(k\right)&\simeq&\left[\log\left(\frac{k}{\sqrt{2}H}\right)+2\frac{m^2}{v^2}\frac{k^2\Mp^2}{H^4} N_\mathrm{c}\right]\Big/\nonumber\\
& &\left(1-2\frac{m^2}{v^2}\frac{k^2\Mp^2}{H^4}\right)\, .
\eea

\subsubsection{Quantitative mode evolution analysis}
 
Now that we have analyzed qualitatively the behavior of the mode function as they cross their ``effective'' Hubble radius, let us move on to massaging the equation for the mode function into a more practical form for calculations. Considering the form (\ref{modefctmassPsi2-phidep}) of writing the last term appearing in $m_u^2(\tau)$, one has
\bea
\label{effectivemass2}
m_u^2(\tau)=\frac{2+15\varepsilon- 3\frac{\lambda\sigma_\chi^2}{H^2}-\frac{12\Mp^2}{v^2} \left\lbrace \mathrm{e}^{\left[ N_c-N\left(\tau\right)\right]\frac{2m^2}{3H^2}}-1\right\rbrace}{\tau^2}\, .\nonumber\\
\eea

Now, since we are interested in the late-time behavior of the mode functions, that is, after they exit their effective Hubble radius, we use an asymptotic approximate solution for the scaling of their amplitude. To do this, let us define a differential equation for an ``effective'' scale factor $\mathfrak{a}$:
\be
	\frac{\mathfrak{a}^{\prime\prime}}{\mathfrak{a}}=m^2_u(\tau)\, .
\ee
We call this quantity the ``effective'' scale factor because it allows us to rewrite the equation for the mode functions in the standard form for a massless field in de-Sitter space:
\be
	u_{\bm{k}}^{\prime\prime}+\left(k^2-\frac{\mathfrak{a}^{\prime\prime}}{\mathfrak{a}}\right)u_{\bm{k}}=0\, .
\ee
Moreover, for a single massless scalar field in de-Sitter space, on super-Hubble scales, the mode functions (call them $z_{\bm{k}}$) scale as the scale factor $a$ if one waits long enough for the decaying mode to become negligible. This means that for small $k^2<\mathrm{min}\left(2/\tau_1^2, 2/\tau_2^2\right)$, one has $a(\tau_1)/a(\tau_2)\approx z_{\bm{k}}(\tau_1)/z_{\bm{k}}(\tau_2)$ (neglecting an overall irrelevant constant phase), provided $\tau_1$ and $\tau_2$ are chosen to be long enough after the Hubble-crossing of the $\bm{k}$-mode (which usually means only a few $e$-folds).

Here we are facing a similar situation. Qualitatively, once the modes $u_{\bm{k}}$ cross their effective Hubble radius (technically a few $e$-folds after the crossing), they scale as the effective scale factor $\mathfrak{a}(\tau)$ (if the modes are under-damped after their effective Hubble-crossing, \ie for the modes such that $k<k_\mathrm{c}$, one simply needs to be somehow more careful about the matching of the sub- and super-$H_{\mathrm{eff}}$ scalings, but the same argument essentially still holds). Since $\mathfrak{a}$ is basically given by the background equation of motion with a non-vanishing $\sigma_ \chi$, one finds that the $u_{\bm{k}}$'s evolve asymptotically as the linearized background after $H_\mathrm{eff}$ crossing. 

This argument provides one with the asymptotic behavior for the evolution of the norm of the super-$H_\mathrm{eff}$ $\delta\psi_{\bm{k}}^{(1)}$ modes:
\bea
	\delta\psi_{\bm{k}}^{(1)}(\tau)&=&
	a^{-1}(\tau)\left\vert u_{\bm{k}}(\tau)\right\vert \nonumber\\
		&\simeq&a^{-1}(\tau)\frac{\mathfrak{a}(\tau)}
		{\mathfrak{a}(\tau_{H_\mathrm{eff}})}
		\left|u_{\bm{k}}(\tau_{H_\mathrm{eff}})\right|\nonumber\\
		&\simeq&a^{-1}(\tau)\frac{\mathfrak{a}(\tau)}
		{\mathfrak{a}(\tau_{H_\mathrm{eff}})}
		\left|u_{\bm{k}}(\tau_i)\right|\, .
\eea
Here, $\tau_{H_\mathrm{eff}}$ is defined as the conformal time at which the $\bm{k}$ mode crosses its effective Hubble radius. As before, an overall irrelevant phase factor is neglected. Also, in the last step we use the fact that before effective Hubble crossing one has $m_u\ll k^2$ and the modes just oscillate with constant amplitude. Then $\left|u_{\bm{k}}(\tau_i)\right|$ can be evaluated in the Bunch-Davies initial vacuum. A more precise calculation would consist in finding the exact sub-Hubble solution, given in terms of Hankel functions of the first kind (once the Bunch-Davies initial conditions are imposed), and evaluating it at effective Hubble-crossing. However, not much accuracy would be gained by doing so. 

This being said, one is only left with the problem of solving the differential equation for the effective scale factor $\mathfrak{a}$ and inverting the relation $k^2=m^2_u\left[\tau_{H_\mathrm{eff}}\left(k\right)\right]$ to obtain $\tau_{H_\mathrm{eff}}\left(k\right)$.

The first problem is an easy one since it just corresponds to solving the linearized background equation of motion. Expressed in terms of the number of $e$-folds, it is given by
\bea
\label{fakeafullevolutioneq}
	& &\frac{\dd^2\mathfrak{a}}{\dd N^2}
	+\frac{\dd \mathfrak{a}}{\dd N}-\mathfrak{a}
	\left( 2+15\varepsilon_1- 3\frac{\lambda\sigma_\chi^2}{H^2} 
	\right.	\nonumber\\& &\qquad\quad \left. 
	- \frac{12\Mp^2}{v^2}  \left\lbrace 
	e^{\left[ N_c-N(\tau)\right]\frac{2m^2}{3H^2}}-1\right\rbrace \right)
	=0\, .\nonumber\\
\eea
As before, $\dd/\dd N\left(\sigma_\chi^2/H^2\right)=1/(4\pi^2)$ and the time variation of the $\sigma_\chi$ term in the above equation is suppressed by a $\lambda$ factor. It can therefore be neglected, allowing the solution to be approximated in terms of Bessel functions of the first and second kind:
\bea
\label{fakeafullevolutionsol}
\mathfrak{a}&=&e^{-\frac{N(\tau)}{2}}\left[C_1\mathrm{J}_{\nu}(x)+C_2\mathrm{Y}_{\nu}(x) \right]\, ,\\
&&\mathrm{where}~~\nu=\frac{3H^2}{m^2}\sqrt{\frac{9}{4}+15\varepsilon_1-3\frac{\lambda\sigma_\chi^2}{H^2}+\frac{12\Mp^2}{v^2}} \nonumber\\
&&\mathrm{and}~~~x=\frac{3H^2}{m^2}\frac{2\sqrt{3}\Mp}{v}e^{(N_c-N)\frac{m^2}{3H^2}}\, ,\nonumber
\eea
where $C_1$ and $C_2$ are integrating constants. To fix them, one first notices that an overall constant in $\mathfrak{a}$ bears no physical meaning, since only the ratio $\mathfrak{a}(\tau)/\mathfrak{a}(\tau_i)$ enters in the quantities to be computed. Therefore one only needs to fix the ratio in which the two independent solutions enter in the mode function. To do so, one notes that in the formal limit $k\tau\rightarrow -\infty$, the positive mode function starts out in the Bunch-Davies vacuum and its evolution deep inside its effective Hubble radius is given by a Hankel function of the first kind. It is therefore natural that its approximate behavior after the crossing of its effective Hubble radius be also mapped to another Hankel function of the first kind. One can therefore choose the constants $C_1$ and $C_2$ so that the solution is written in the form of a $H_\nu(x)$ function. 

To give another, maybe more convincing, argument to fix $C_1$ and $C_2$, we note that if $k<k_{\mathrm{v}}$, that is, if the mode is still under-damped and continues its oscillations outside its Hubble radius for a (more or less long) time before freezing out at $N=N_{\mathrm{v}}$, the requirement of having oscillations damped by the factor of $e^{-N/2}$ in Eq.~(\ref{fakeafullevolutionsol}) basically fixes the constants to $C_1=1$ and $C_2=i$, up to an overall irrelevant constant phase. This is precisely the choice that allows to recover the Hankel function of the first kind discussed above.

If on the contrary, $k_{\mathrm{v}}<k<k_\uc$, \ie if the $\bm{k}$-mode of interest crosses its effective Hubble radius late enough so that it is over-damped and freezes out immediately after the crossing, then, provided the mapping is done a few $e$-folds after the crossing, the decaying mode $J_\nu$ of the fundamental solution (\ref{fakeafullevolutionsol}) has completely decayed and the positive mode-function is exclusively mapped to the growing mode $Y_\nu$ to great accuracy. This corresponds to only using the Bessel function of the second kind $Y_{\nu}$ as a solution, which is also approximately what using a Hankel function of the first kind would mean in the relevant range of values for $\nu$ and $x$. 

However it might be cumbersome to work in terms of Bessel or Hankel functions, mainly because for the regime of parameters $\nu$ and $x$ one is interested in, none of the asymptotic forms of theses functions are good approximations when the mode function freezes out and is mapped to the growing mode. Indeed, the small-argument form holds if $x\ll\sqrt{\nu}$, which here is not the case since one works under vacuum domination, and the large-argument expansion is valid provided $x\gg\nu^2$, which is not the case either, again because of vacuum domination.

It is therefore useful to note that, since $\frac{2m^2}{3H^2}\ll 1$, one can Taylor expand to first order the exponential in Eq.~(\ref{fakeafullevolutioneq}), in order to obtain a simpler differential equation which can be solved in terms of Airy functions:
\bea
\label{fakeaTayloredEvolutionEqn}
	\frac{\dd^2\mathfrak{a}}{\dd N^2}
	+\frac{\dd\mathfrak{a}}{\dd N}-\mathfrak{a}
	\left\lbrace 2+15\varepsilon- 3\frac{\lambda\sigma_\chi^2}{H^2}
	\right.\qquad \qquad\qquad\qquad \\ \left.
	-\frac{8\Mp^2m^2}{v^2H^2}\left[N_\uc-N(\tau)\right]\right\rbrace
	=0 \, ,\nonumber
\eea
which is solved by:
\be
	\mathfrak{a}=e^{-\frac{N}{2}}\left[
	\mathrm{Ai}\left( x\right)C_1+\mathrm{Bi}\left(x\right)C_2\right]\, ,
\ee
with
\bea 
	x&=&\left(\frac{v^6\lambda}{96m^2\Mp^4}\right)^{\frac{2}{3}}
	\left[\frac{96m^2\Mp^4}{v^6\lambda}(N-N_c)
	\qquad\qquad\qquad\qquad\right.\nonumber\\ 
	& & \qquad\qquad\qquad\qquad \left.
	-3\frac{\lambda\sigma_\chi^2}{H^2}
	+15\varepsilon_1+\frac{9}{4} \right]\, ,
\eea
where $C_1$ and $C_2$ are integration constants which are not necessarily the same as before. For the $\bm{k}$-modes such that $k<k_\mathrm{v}$, which are under-damped when they cross their effective Hubble radius, one has $x\ll 0$ and the asymptotic forms of the Airy functions for large and negative arguments in terms of sine and cosine can be used. Since oscillations are expected, one can choose $C_1=i$ and $C_2=1$. Then, deep inside the valley when these modes cross their effective Hubble radius, one obtains:
\be 
	\mathfrak{a}\underset{x\rightarrow -\infty}{=}
	\frac{e^{-N/2}}{\sqrt{\pi}}|x|^{-1/4}
	e^{i\frac{2}{3}|x|^{3/2}+\frac{i}{4}\pi}\, .
\ee
The modes that freeze out in that limit evolve according to
\bea
	\delta\psi_{\bm{k}}^{(1)}&\approx& 
	e^{\frac{1}{2}(N_{H_\mathrm{eff}}-3N)}
	\left\vert\frac{x(N)}{x(N_{H_\mathrm{eff}})}\right\vert^{-\frac{1}{4}}
	\frac{e^{i\frac{2}{3}|x(N)|^{\frac{3}{2}}}} 
	{e^{i\frac{2}{3}|x(N_{H_\mathrm{eff}})|^{\frac{3}{2}}}} 
	\nonumber\\
	&&\qquad \times \left| u_{\bm{k}} (N_{H_\mathrm{eff}}) \right|\\
	&\approx& e^{\frac{1}{2}\left[N_{H_\mathrm{eff}}-3N\right]
	-i\frac{2}{3}\left[|x(N_{H_\mathrm{eff}})|^{\frac{3}{2}}- 
	|x(N)|^{\frac{3}{2}}\right]}
	\nonumber\\
	\label{solutionunderdampedearly}
	&& \qquad \times\left\vert\frac{x({N_{H_\mathrm{eff}}})}
	{x(N)}\right\vert^{\frac{1}{4}}\frac{1}{\sqrt{2k}}\\
	&\mathrm{for}&x(N),~x(N_{H_\mathrm{eff}}) 
	\ll0\quad(\mathrm{and}~k^2<m_u^2)\, .\nonumber
\eea
In the first line of this equation one can see the previously mentioned oscillations, which were expected to be found since in that limit one has $m_u^2(N)<0$. One also finds the decay factor $e^{-3N/2}$. Recall that in order to express $N_{H_\mathrm{eff}}$ in terms of $k$ in that regime, one needs to solve $|k_{H_\mathrm{eff}}^2|=|m_u^2(N_{H_\mathrm{eff}})|$.

If $k>k_{\mathrm{v}}$, $m_u^2(N_{H_\mathrm{eff}})$ becomes positive, the oscillations cease and the modes freeze out. To see this, one can equivalently examine $x$, which becomes positive as $x\rightarrow \mathrm{constant} \times [9/4+15\varepsilon-3\lambda\sigma_\chi^2/(H^2)]^-$, and the above approximation for the Airy functions breaks down. However, to find how the behavior of the modes $k<k_\mathrm{v}$ changes when $N>N_{\mathrm{v}}$, and to derive the behavior of the modes with $k_\mathrm{v}<k<k_\mathrm{c}$ which cross their effective Hubble radius in that limit, one can assume a long waterfall to take place (which we recall to be necessary in order to evade the blue tilt problem) and suppose $v^6\lambda\gg m^2\Mp^4$ [see Eq.(\ref{eq:efoldswater})], to use the large argument expansion of the Airy functions and proceed as above. In this limit, one obtains a growing mode and a decaying mode, and keeping only the former in the asymptotic solution, one gets
\bea
	\mathfrak{a}&\underset{x\rightarrow +\infty}{=}&
	\frac{e^{-\frac{N}{2}}}{\sqrt{\pi}}|x|^{-1/4}e^{\frac{2}{3}x^{3/2}}\, .
\eea
Using this asymptotic expression, one obtains, for the under-damped modes $k<k_{\mathrm{v}}$, once frozen out (for $N>N_{\mathrm{v}}$),
\bea
\label{solutionunderdampedlate}
	\delta\psi_{\bm{k}}^{(1)} \simeq e^{\frac{1}{2}
	(N_{H_\mathrm{eff}}-3N)}\left\vert\frac{x(N_{H_\mathrm{eff}})}
	{x(N)}\right\vert^{\frac{1}{4}}\frac{e^{\frac{2}{3}x(N)^{\frac{3}{2}}}}
	{e^{\frac{2}{3}x(N_\mathrm{v})^{\frac{3}{2}}}}\frac{1}{\sqrt{2k}}
	\nonumber \\ \mathrm{for~} x(N)\gg 0,~x(N_{H_\mathrm{eff}}) \ll0
	\quad(\mathrm{and}~k^2<m_u^2)\, ,
\eea
and where a constant irrelevant phase factor is neglected. For modes $k>k_{\mathrm{v}}$ freezing out in that regime [for which $x(N_{H_\mathrm{eff}})>0$], one has
\bea
	\delta\psi_{\bm{k}}^{(1)}&\simeq & e^{\frac{1}{2}
	(N_{H_\mathrm{eff}}-3N)}\left\vert\frac{x(N)}
	{x(N_{N_{H_\mathrm{eff}}})}\right\vert^
	{-\frac{1}{4}}\frac{e^{\frac{2}{3}x(N)^{\frac{3}{2}}}} 
	{e^{\frac{2}{3}x(N_{H_\mathrm{eff}})^{\frac{3}{2}}}} \nonumber\\
	&&\qquad \times \left| u_{\bm{k}} (N_{H_\mathrm{eff}}) \right|\\
	&\simeq & e^{\frac{1}{2}\left[N_{H_\mathrm{eff}}-3N\right]
	+\frac{2}{3}\left[x(N)^{\frac{3}{2}}- x(N_{H_\mathrm{eff}})^{\frac{2}{3}}
	\right]}\nonumber\\	\label{solutionoverdamped}&&\qquad \times
	\left\vert\frac{x(N_{N_{H_\mathrm{eff}}})}{x(N)}\right\vert^
	{\frac{1}{4}}\frac{1}{\sqrt{2k}}\\
	&\mathrm{for}&x(N),x(N_{H_\mathrm{eff}})\gg 0
	\quad(\mathrm{and}~k^2<m_u^2)\, .\nonumber
\eea

The formulae derived above for the amplitude of the first order perturbations in the $\psi$ direction are collected  in Appendix \ref{app:deltapsiFormula}, see Eqs.~(\ref{PsimodesAnalBegin}-\ref{PsimodesAnalEnd}), for practical convenience. 

\begin{figure*}
\begin{center}
\includegraphics[width=8.5cm]{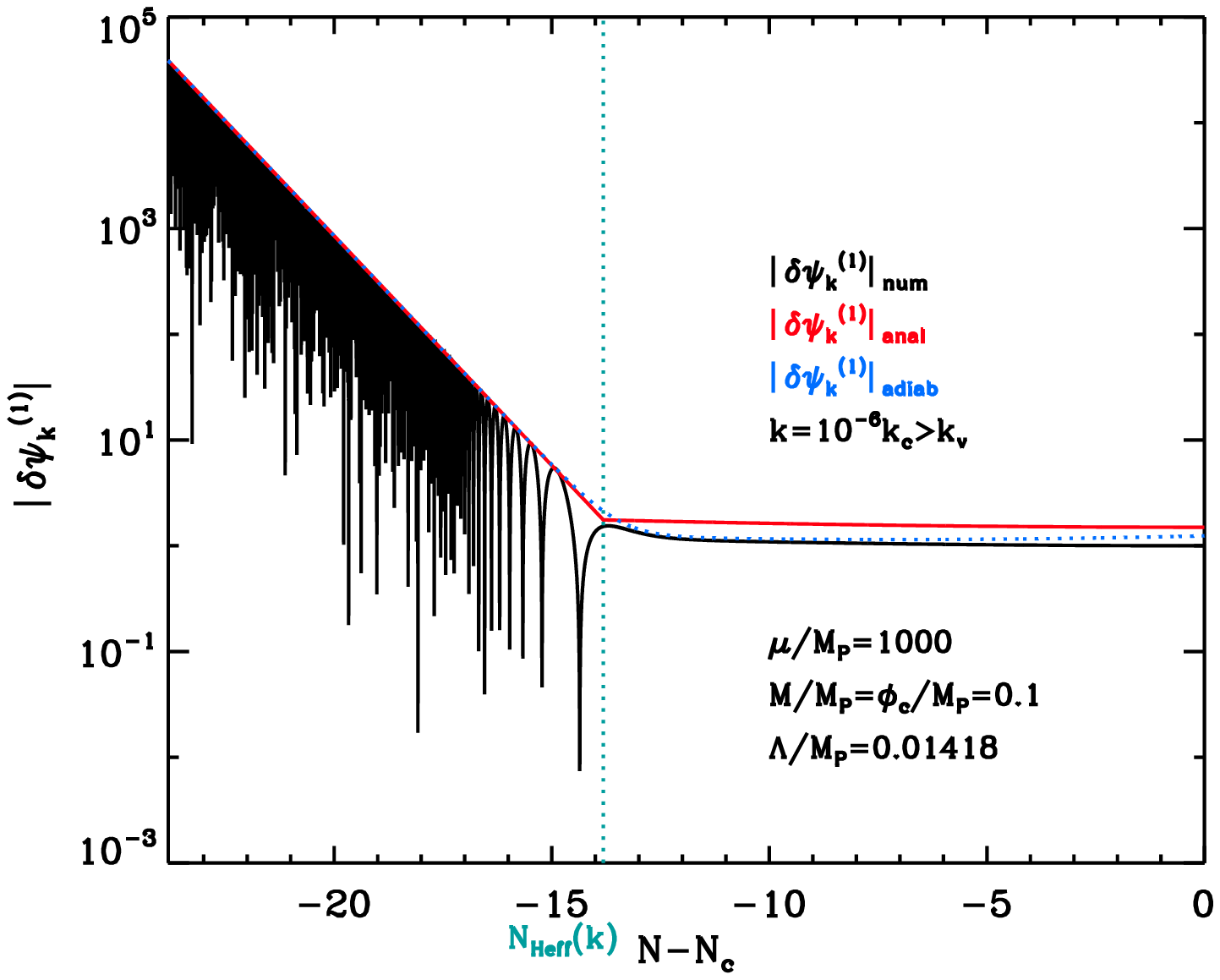}
\includegraphics[width=8.5cm]{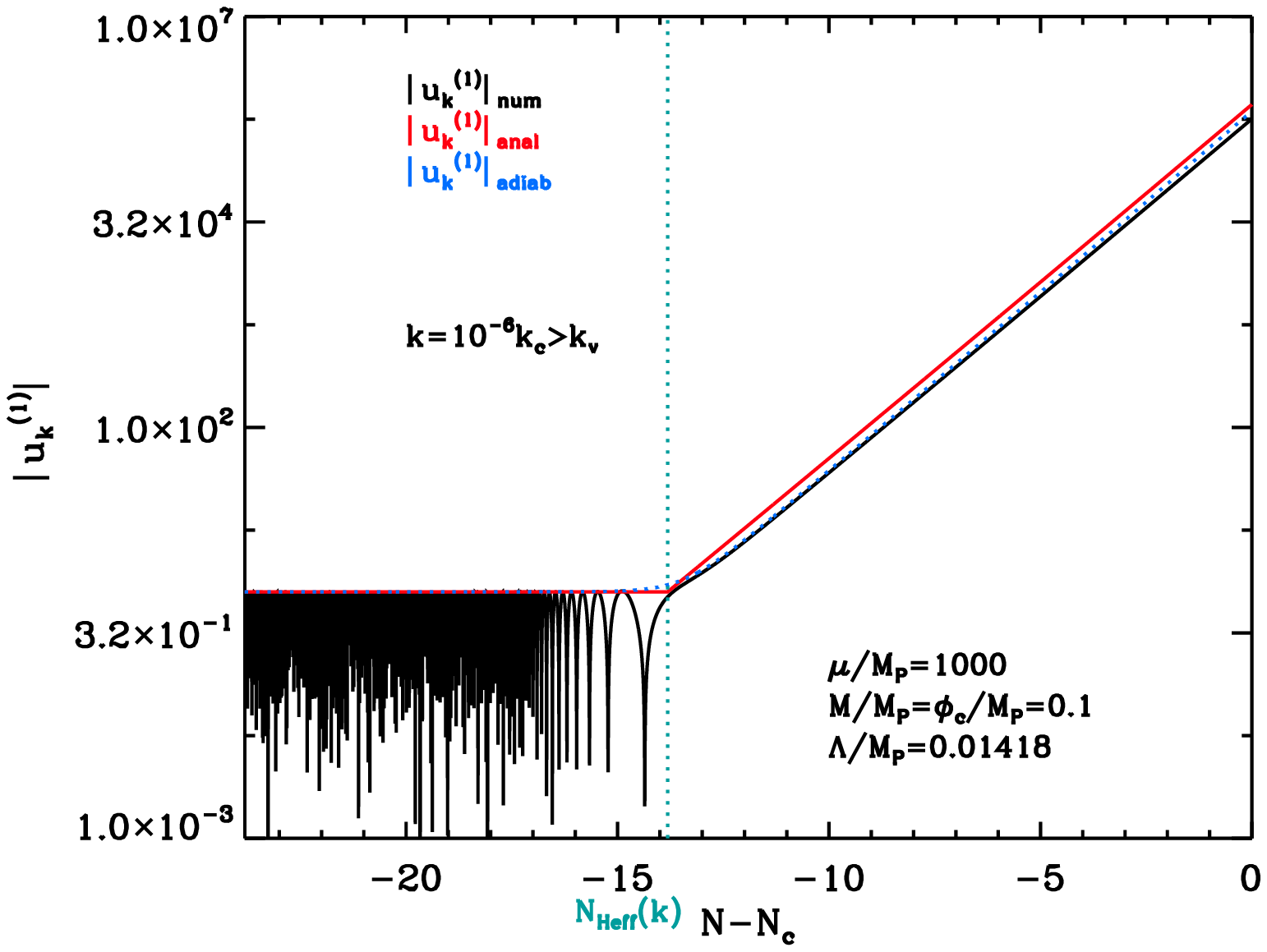}
\includegraphics[width=8.5cm]{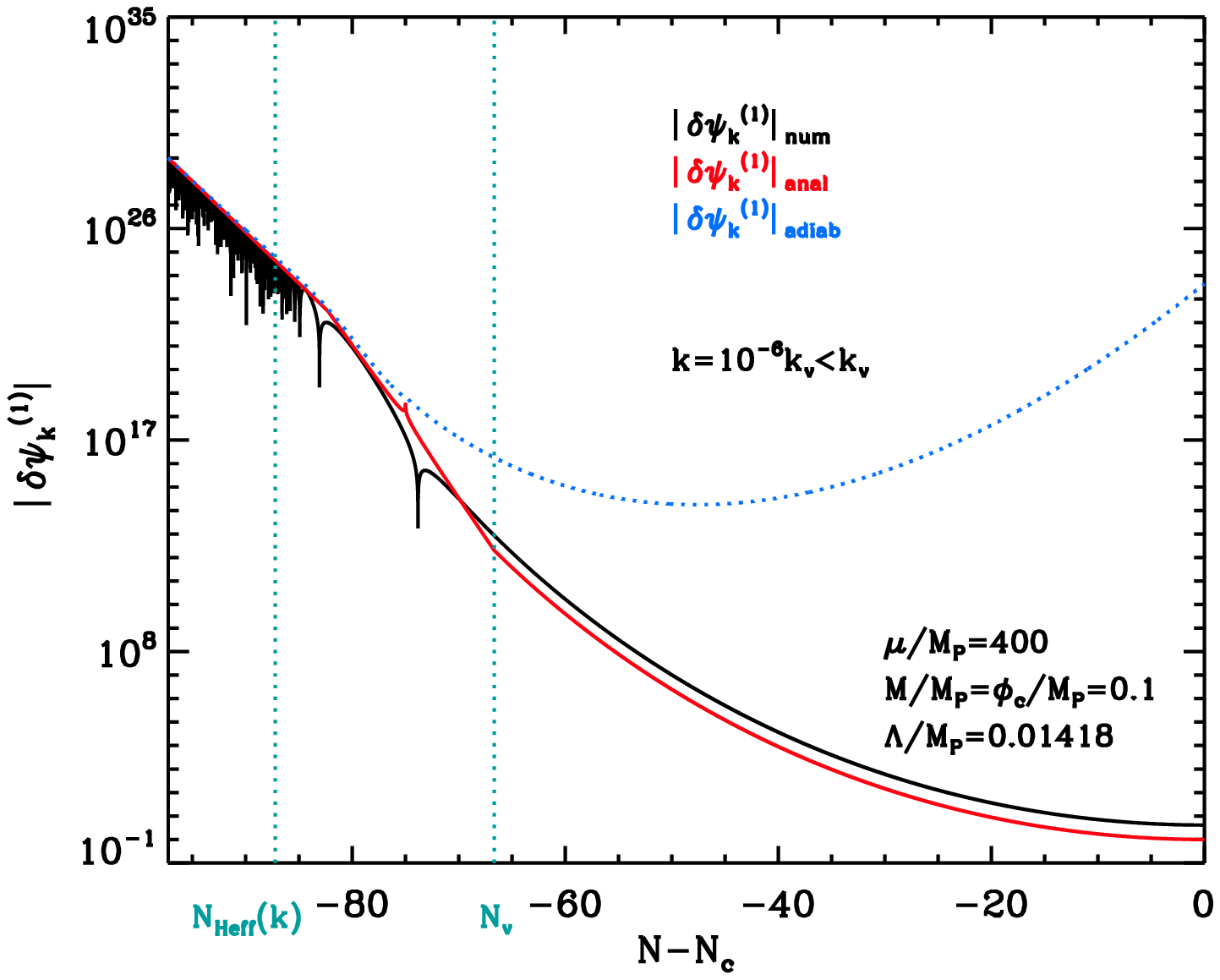}
\includegraphics[width=8.5cm]{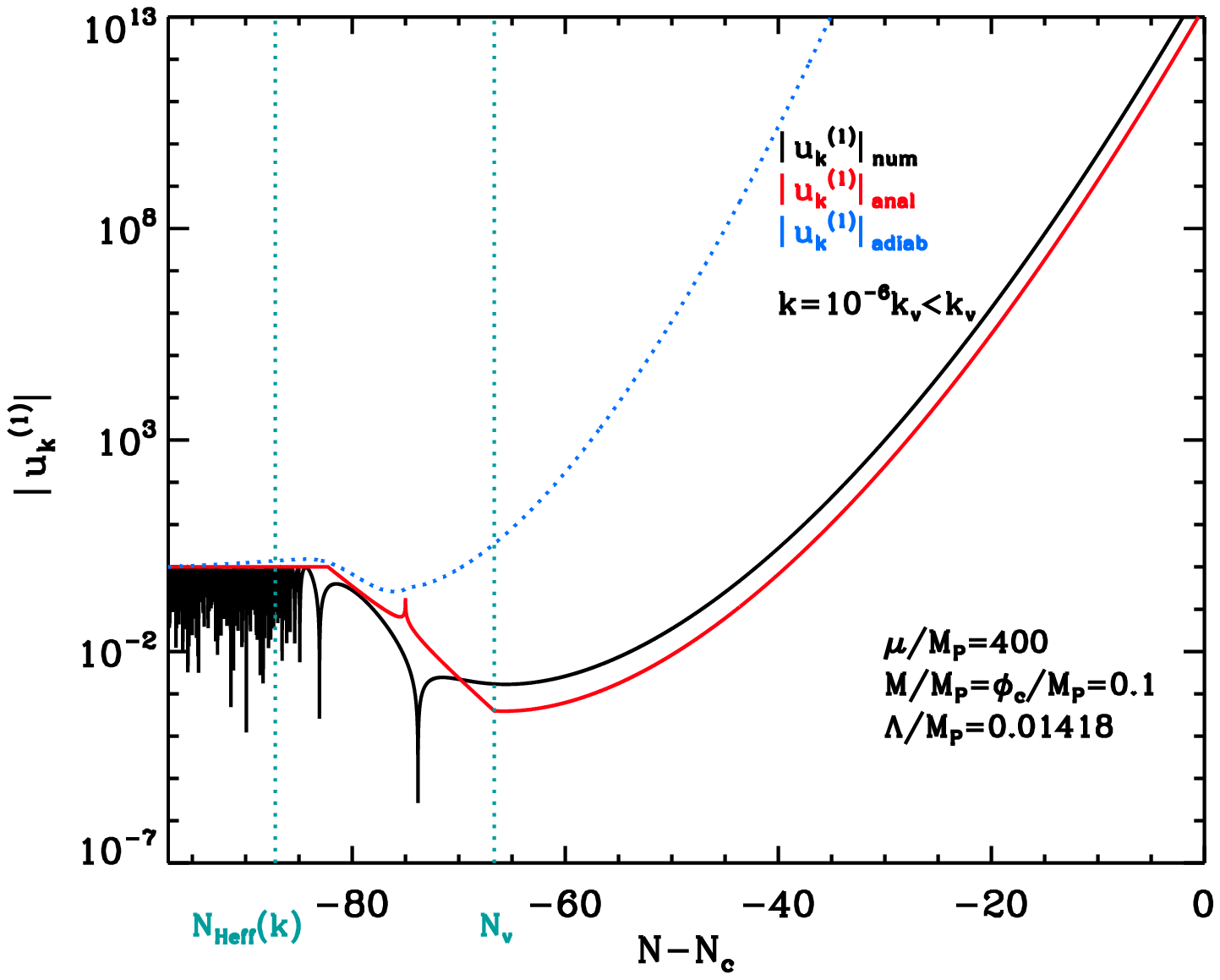}
\caption{Time evolution of the first order perturbation amplitudes of the waterfall field $\vert\delta\psi_{\bm{k}}^{(1)}\vert$ (left panels) and its scaled counterpart $u_{\bm{k}}$ (right panels) in the cases $k>k_{\mathrm{v}}$ (top panels) and $k<k_{\mathrm{v}}$ (bottom panels). The black solid lines are numerical results from an exact integration of Eq.~(\ref{ukgenericmodefunction}). The red solid lines represent the analytical approximated results Eqs.~(\ref{PsimodesAnalBegin}-\ref{PsimodesAnalEnd}). The blue dotted lines represent the adiabatic solution Eq.~(\ref{psimodesAdiab}). When $k>k_{\mathrm{v}}$ the modes evolve in the standard well known way, $\vert u_{\bm{k}}\vert$ being constant on sub-Hubble scales and $\vert\delta\psi_{\bm{k}}^{(1)}\vert$ being constant on super-Hubble scales. When $k<k_{\mathrm{v}}$ however, there is a intermediate phase $N_{H_{\mathrm{eff}}}(k)<N<N_{\mathrm{v}}$ during which the adiabatic evolution of the effective mass of the waterfall breaks down and the fluctuations are over-damped.}
\label{fig:psimodes}
\end{center}
\end{figure*}

Before proceeding, since several approximations have been performed, it seems useful to first check their validity by comparing them with the full numerical integrations of Eq.~(\ref{ukgenericmodefunction}). We also check the validity of the commonly-used so-called {\it adiabatic approximation}. This scheme is defined as follows: since the inflaton field is slowly rolling down the bottom of the valley, the effective mass for the waterfall field $m_\psi$, defined as $m_u^2\equiv (2-m_\psi^2/H^2)/\tau^2$ in Eq.~(\ref{effectivemass1})  ($m_u$ is sketched in Fig.~\ref{fig:psimasssketch}), is varying slowly and therefore its time dependence can be neglected. Hence, when solving Eq.~(\ref{ukgenericmodefunction}), the usual constant-mass mode function solution
\begin{equation}
u_{\bm{k}}\simeq -i\mathrm{e}^{i\left(\nu+\frac{1}{2}\right)\frac\pi2}
\frac{\sqrt{\pi}}{2}\left(-\tau\right)^{1/2}H_\nu^{(1)}\left(-k\tau\right)\, ,
\label{psimodesAdiab}
\end{equation}
can be used with $\nu$ now given by the time varying quantity $\nu\equiv\sqrt{9/4-m_\psi^2/H^2}$, which is complex when $N<N_{\mathrm{v}}$. This approximate solution is referred to as the adiabatic one since it is derived under the approximation of a slowly-varying mass. We check its validity in Fig.~\ref{fig:psimodes}, where results from an exact integration of Eq.~(\ref{ukgenericmodefunction}) are compared to the analytical approximations~(\ref{PsimodesAnalBegin}-\ref{PsimodesAnalEnd}) and to the adiabatic solution Eq.~(\ref{psimodesAdiab}).

Let us comment on what has been obtained. The case $k>k_{\mathrm{v}}$ is similar to the standard well known massless case, where $\vert u_{\bm{k}}\vert$ is constant on sub-Hubble scales and $\vert\delta\psi_{\bm{k}}^{(1)}\vert$ is constant on super-Hubble scales. The matching between the analytical expressions~(\ref{PsimodesAnalBegin}-\ref{PsimodesAnalEnd}) and the numerical solution is excellent. The adiabatic approximation also holds during the whole evolution of such modes.

If $k<k_{\mathrm{v}}$, there is an intermediate regime when $N_{H_{\mathrm{eff}}}(k)<N<N_{\mathrm{v}}$ where the field fluctuations are over-damped and oscillations continue to take place. As noticed in Fig.~\ref{fig:psimasssketch}, during that phase, at some point, such modes experience a short period during which $k^2$ dominates over $m_u^2$ again. Since this period is very short in time, it was not taken into account when deriving the analytical expressions Eqs.~(\ref{PsimodesAnalBegin}-\ref{PsimodesAnalEnd}) and checking this assumption was postponed to later. One can now check that it indeed leads to rather reliable expressions. However, this short phase of rapid evolution of $m_u^2$ obviously breaks the adiabatic approximation and one can indeed see that the adiabatic formula stops being valid at this point. The subsequent evolution is therefore also different from the one one expects under the adiabatic approximation. When $N>N_{\mathrm{v}}$, the field fluctuations continue to experience over-damping.

The validity of the analytical expressions derived above is thus confirmed, as well as the schematic description previously sketched, and the adiabatic method is shown not to be valid for the modes such that $k<k_{\mathrm{v}}$, when $N$ approaches $N_{\mathrm{v}}$ and afterward.

\section{Calculation of $\sigma_\chi$ and the back-reaction problem}
\label{sec:sigmapsi}

So far we have calculated to leading order in the slow-roll parameters the amplitude of the linearized quantum fluctuations in both fields in the presence of a shifted background. These are the $\mathcal{O}(\hbar^2)$ mode functions defined in Eqs.~(\ref{modefunctiondefphi}) and (\ref{modefunctiondefpsi}) up to leading order in slow-roll which enter in the bath fields propagators evaluated at the time when a given mode of the quantum fields joins the coarse-grained fields. We can therefore use these results to directly calculate a shifted classical noise for equations (\ref{Langphi}) and (\ref{Langpsi}), which will now be valid to $\mathcal{O}(\hbar^2)$ and to leading order in slow-roll.

This higher accuracy calculations does bear some importance. Indeed, the typical deviation in the waterfall direction acquired during the valley phase sets typical initial conditions for the subsequent waterfall phase, hence determining how many $e$-folds this tachyonic period should last. One should therefore calculate ${\sigma_\chi}_\uc$ as accurately as possible.

A first estimate was given in section~\ref{sec:masslessDSdispersions} using the standard massless de-Sitter solutions for the modes $\delta\phi^{(1)}$ and $\delta\chi^{(1)}$ to calculate the amplitude of the noises, see Eq.~(\ref{sigmapsi}). This was a first step towards a more accurate calculation, carried out mainly to obtain qualitative results. We now wish to include the higher-accuracy noises derived from the mode amplitude results of the previous subsection, and study how this impacts on the stochastic dispersions of the coarse grained fields, which we do here, and on the statistics of the inflaton perturbations, which we do in the next section. 

For the coarse-grained inflaton field, $\varphi$, we expect the solution of Eq.~(\ref{Langphi}) including higher noise accuracy to closely follow the noiseless, classical solution. This is because we already assumed the values of the potential parameters to be such that the dynamics of the inflation in the valley phase is dominated by its classical drift at $\mathcal{O}(\hbar)$. We can convince ourselves that this assumption is preserved at $\mathcal{O}(\hbar^2)$ and to leading order in the two first slow-roll parameter by looking at the corrected inflaton noise auto-correlation, which shows a suppressed correction compared to its $\mathcal{O}(\hbar, \epsilon_1^0, \varepsilon_2^0)$ value:
\bea  
\label{theshiftedinflatonnoisespectrum}
\left\langle\xi_\phi\left(N\right)\xi_\phi\left(N^\prime\right)\right\rangle=
\frac{H^4}{4\pi^2}\delta\left(N-N^\prime\right)
\times\nonumber\\
\left[1+\frac{2}{3}\left(\frac{m^2+g\sigma_\chi^2}{H^2} -9\varepsilon_1\right) 
\left(\ln 2\epsilon +\gamma-2\right)\right]\, .
\eea
Here $\gamma\simeq 0.577$ is the Euler-Mascheroni constant, and recall that $\varepsilon_2\approx-\frac{2}{3}\frac{m^2}{H^2}$. The correction to the de-Sitter massless formula~(\ref{noisecorrphideSitter}) is indeed small for a light inflaton field, and one does not expect important effects on the background trajectory coming from a better calculation of $\sigma_\phi$. Important effects concerning the inflaton, however, are to be expected when it comes to the statistics of the fluctuations of the coarse-grained field, and will be calculated in the next section.

As a final remark concerning the background inflaton coarse-grained field, it is interesting to remember that in spite of the fact that the condition $\epsilon\ll 1$ was required in order to only collect the squeezed super-Hubble modes in the coarse-grained part of the field, the splitting parameter $\epsilon$ cannot be arbitrarily small if one wants the deviations from the free massless case to remain small. More precisely, from the previous equation, one can see that the condition $\exp(-H^2/m^2)\ll\epsilon\ll 1$ should be imposed. This is exactly the condition that was obtained by Starobinsky and Yokoyama in their first paper \cite{Starobinsky:1994bd} on the subject [see Eq.~(81) there], requiring $\epsilon$-independent results for the two point equilibrium correlation function of test scalar fields in de-Sitter. Here we make the origins of such a condition rather clear.

On the other hand, the waterfall field is significantly massive far enough in the valley. Therefore, important effects on its dispersion coming from a higher-order calculation of the noise sourcing its coarse-grained evolution are expected to arise in this region. We shall investigate this question in detail in what follows. Whether these effects can lie in the observational window or not is also a question which shall be answered.

Concretely, the higher-order white Gaussian noise $\xi_\psi\left(N\right)$ sourcing the Langevin equation~(\ref{Langpsi}) for the coarse-grained waterfall field is given by  
\be
\left\langle \xi_\psi\left(N\right) \xi_\psi\left(N^\prime\right) \right\rangle \propto \left\vert \delta\psi_{\bm{k}}^{(1)} \right\vert_{k=\epsilon aH}^2 \delta\left(N-N^\prime\right) \, ,
\ee
with $\vert\delta\psi_{\bm{k}}^{(1)}\vert$ now evaluated using Eqs.~(\ref{PsimodesAnalBegin}-\ref{PsimodesAnalEnd}).

Before proceeding to this evaluation, a verification is in order. In the computational program described above, one should remember that $vert \delta\psi^{(1)}_{\bm{k}}\vert$ takes different forms depending on whether $N\lessgtr N_\mathrm{v},N_{H_\mathrm{eff}}$ and one needs to know which piece of the function should be used. Furthermore, $\epsilon$ is usually taken to be such that $\epsilon\ll 1$ in order to keep only the super-Hubble highly squeezed modes in the coarse grained field (squeezing being the condition for classical behavior, see Refs.~\cite{Guth:1985ya,Polarski:1995jg,Kiefer:1998qe}).

However, here, the effective Hubble radius $H_\mathrm{eff}^{-1}$ intervenes rather than the Hubble radius itself and therefore one first needs to be sure that no modification to the standard picture arises from this fact. In Ref.~\cite{Mijic:1997mt}, the original analysis of Guth and Pi \cite{Guth:1985ya} is generalized to heavy fields and it is found that there is no emergence of classical correlations for $\nu^2<0$ (recall that $\nu_\psi^2=9/4-m_\psi^2/H^2$). Such classical correlations, usually obtained through turning of quantum oscillators upside-down or by rapid squeezing of upside-right oscillators, are a key point of the stochastic inflation formalism as they enable to treat the dynamics of large wavelength fluctuations as following a stochastic classical evolution.

For our purpose, it means that when $N<N_{\mathrm{v}}$, the stochastic equations driving the evolution of the coarse grained field are questionable, and that a full field theoretic approach should be used instead. Therefore in the following, one should be careful when interpreting the results derived for $N<N_{\mathrm{v}}$.

Recalling that 
\be 
N_\uc-N_{\mathrm{v}}=\lambda v^6/\left(48m^2\Mp^4\right)\propto\left(N_{\mathrm{end}}-N_\uc\right)^2 
\ee
[see Eq.~(\ref{eq:efoldswater})], this means that such a ``problematic'' period happens long before the critical point if the waterfall  lasts for a long number of $e$-folds. In this case, it does not affect fluctuations in the observational window. In the opposite case (short-lived waterfall), it is on stage almost until the critical point crossing right before the end of inflation, and the interpretation of the stochastic formalism is problematic. Here we only study long-lived waterfall scenarios.

\subsection{Quasi-Stationary Approximation}

\begin{figure}
\begin{center}
\includegraphics[width=9cm]{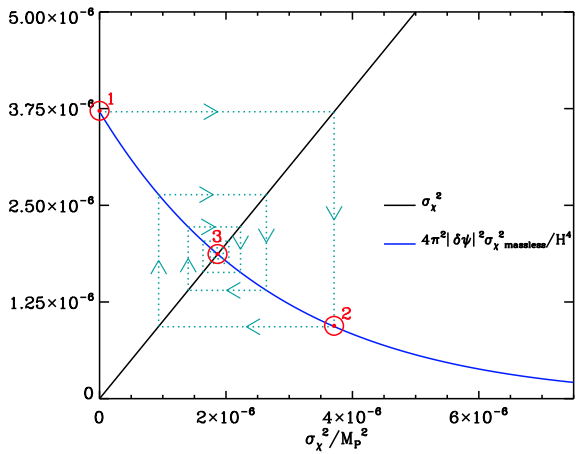}
\caption{Right hand side of Eq.~(\ref{eq:sigmapsiImplicit}), namely $4\pi^2\vert\delta\psi^{(1)}\vert^2{\sigma^2_{\chi\, \mathrm{massless}}}/H^4$ (blue solid line), normalized to $\Mp^2$, as a function of $\sigma_\chi^2$, for $v=0.1503\Mp$, $m=7\times 10^{-5}\Mp$, $g=\sqrt{\lambda}=0.885$, computed one $e$-fold before crossing the critical point. These values may not be physical (especially for $g$ and $\lambda$) but they have been chosen for display convenience. The black solid line is the left hand side of Eq.~(\ref{eq:sigmapsiImplicit}), namely $\sigma_\chi^2$, normalized to $\Mp^2$, so that the solution of Eq.~(\ref{eq:sigmapsiImplicit}) lies at the intersection of these two lines. The green dotted lines and the arrows indicate how an iterative (perturbative) process leads to this solution, hence showing that the ``classical'' guess lies in the attraction basin of the solution. The meaning of the red circles and the associated labels $1$, $2$ and $3$ is detailed in the text body.}
\label{fig:ImplicitSigmaPsi}
\end{center}
\end{figure}

Let us now turn to the concrete calculation of $\sigma_\chi$. One first notices that Eq.~(\ref{eq:chistoSol}) is completely general and is correct whatever the amplitude of the $\xi_\psi$ noise is. However, Eq.~(\ref{twopointscorrfctPsi}) makes use of the specific de-Sitter massless statistics (\ref{noisecorrpsideSitter}) and the lower incomplete gamma function solution arises when the amplitude of the noise $\langle\xi_\psi^2\rangle$ is time independent. It is not true in general. However, the relaxation time of the statistical distribution~(\ref{eq:chistoSol}) is extremely small since $\lambda v^4/(m \Mp^2)\gg 1$, which means in practical terms that the mass of $\Psi$ decreases so slowly that at each given time, the $\chi$ distribution swiftly acquires its ``stationary'' local dispersion. This kind of adiabatic scheme should not be confused with the adiabatic approximation mentioned in section~\ref{sec:deltapsi} in the calculation of $\delta\psi^{(1)}$. The former describes quasi-stationary stochastic distributions while the latter relies on fluctuation modes crossing the relevant scales faster than their mass typical variation times. This is why to avoid confusion, we may refer to the former as the ``quasi-stationary'' frame in what follows. Under this quasi-stationary approximation, on has $ \sigma_\chi^2/\left.\sigma_\chi^2\right\vert_{\mathrm{massless}} \simeq \langle\xi_\psi^2\rangle / \langle\xi_\psi^2\rangle_{\mathrm{massless}} = \vert \delta\psi^{(1)} \vert^2 / \vert \delta\psi^{(1)} \vert^2_{\mathrm{massless}}$, so that one has 
\be
\label{eq:sigmapsiImplicit}
\sigma_\chi^2\simeq
\frac{\left\vert\delta\psi^{(1)}\right\vert^2}{H^4/(4\pi^2)}
\left.\sigma_\chi^2\right\vert_{\mathrm{massless}}\, ,
\ee
where $\left.\sigma_\chi^2\right\vert_{\mathrm{massless}}$ is given by Eq.~(\ref{sigmapsi}).

It is of particular interest to notice that Eq.~(\ref{eq:sigmapsiImplicit}) is actually an implicit relation involving $\sigma_\chi$, since $\left\vert\delta\psi^{(1)}\right\vert$ involves $\sigma_\chi$ itself [see Eqs.~(\ref{PsimodesAnalBegin}-\ref{PsimodesAnalEnd})]. In some sense, the whole recursive strategy presented in section~\ref{sec:RecursiveStrategy} is now summarized in a single implicit equation for $\sigma_\chi$. The situation is summarized in Fig.~\ref{fig:ImplicitSigmaPsi}, where the left hand side and the right hand side of Eq.~(\ref{eq:sigmapsiImplicit}) are displayed, as a function of $\sigma_\chi^2$. The solution of Eq.~(\ref{eq:sigmapsiImplicit}) lies at the intersection of these two curves, the location of which can be calculated using a recursive scheme which exactly translates the one presented in section~\ref{sec:RecursiveStrategy}. The red circle labeled ``$1$'' in Fig.~\ref{fig:ImplicitSigmaPsi} represents the solution of Eq.~(\ref{eq:sigmapsiImplicit}) when setting $\sigma_\chi=0$ in the right hand side. This is the solution calculated in section~\ref{sec:masslessDSdispersions} (where one has also neglected the mass of $\Psi$). This corresponds to evolving the perturbations $\delta\psi^{(1)}$ on a ``classical'' unshifted background. Then one can source the equation of motion for these perturbations with a background shifted by the value of $\sigma_\chi$ just calculated. This new solution is represented by the red point labeled ``$2$'' in Fig.~\ref{fig:ImplicitSigmaPsi}. This iterative procedure can be continued until obtaining the exact solution labeled by the red circle ``$3$''. 

One can remark that the ``classical'' guess (labeled ``$1$'') lies in the attraction basin of the exact solution (labeled ``$3$''). This is an indication that the perturbative expansion is under control, since at each step, one gets closer to the exact solution and decreases the absolute value of its displacement. This is a direct consequence of the fact that the right hand side of Eq.~(\ref{eq:sigmapsiImplicit}) is a decreasing function of $\sigma_\chi$, which is always true since as $\sigma_\chi$ increases, the mass ``seen'' by the perturbations $\delta\psi^{(1)}$ increases, hence the amplitude of the noise decreases, and so does the resulting $\sigma_\psi$. However, it may be not the case during the waterfall, where this mass becomes more negative as $\sigma_\chi$ increases, rendering the amplitude of the noise more important. This signals a tachyonic breakdown of the perturbative expansion which indicates that the model may face serious issues when carefully studied in the waterfall (especially if this phase is long). We will come back to this point later, explaining how the waterfall start can be delayed. 

\begin{figure}
\begin{center}
\includegraphics[width=9cm]{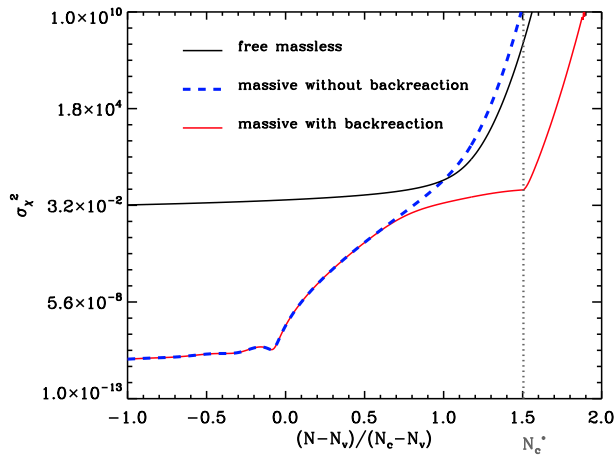}
\caption{Stochastic dispersion in the $\chi$ direction $\sigma_\chi^2$, rescaled by its value~(\ref{eq:sigmapsic}) at the critical point in the free massless case, as a function of time labeled by $(N-N_{\mathrm{v}})/(N_\uc-N_{ \mathrm{c}})$ (which is $1$ at the critical point and $0$ at the point $N=N_{\mathrm{v}}$). The black solid line represents the free massless result~(\ref{sigmapsi}). The blue dashed line takes the mass of $\Psi$ into account but does not include back-reaction. Technically, it corresponds to the right hand side of Eq.~(\ref{eq:sigmapsiImplicit}) setting $\sigma_\chi=0$, \ie the value at the point labeled ``$1$'' in Fig.~\ref{fig:ImplicitSigmaPsi}. The red solid line represents the exact solution of Eq.~(\ref{eq:sigmapsiImplicit}), \ie the value at the point labeled ``$3$'' in Fig.~\ref{fig:ImplicitSigmaPsi}. The parameter values used are $v=0.1503\,\Mp$, $m=2.24\times 10^{-4}\,\Mp$, $g=\sqrt{\lambda}=4.2$. These values may not be physical (especially for $g$ and $\lambda$) but they have been chosen for display convenience. The grey dotted line represents the value of $N_\uc^\star$ defined in Eq.~(\ref{eq:Ncstar}) (see text body).}
\label{fig:SigmaPsi}
\end{center}
\end{figure}

Let us now see how these different estimations of $\sigma_\chi$ evolve in time. In Fig.~\ref{fig:SigmaPsi} are displayed the free massless result~(\ref{sigmapsi}), the result of a calculation taking into account the mass of $\Psi$ but no back-reaction (corresponding to the point labeled ``$1$'' in Fig.~\ref{fig:ImplicitSigmaPsi}), and the exact solution of Eq.~(\ref{eq:sigmapsiImplicit}) (corresponding to the point labeled ``$3$'' in Fig.~\ref{fig:ImplicitSigmaPsi}), as a function of time. When $N\ll N_\uc$ (remember that $N<N_{\mathrm{v}}$ is not obvious to interpret), the inclusion of the mass of $\Psi$ significantly decreases the value obtained for $\sigma_\chi$, since a positive mass better confines the distribution for $\chi$. In this regime $\sigma_\chi$ remains small and the inclusion of back-reaction does not alter much the result. As the system gets closer to the critical point, $\sigma_\psi$ increases and a discrepancy due to back-reaction starts to be visible, which decreases the actual value of $\sigma_\chi$ (in agreement with what is noticed in Fig.~\ref{fig:ImplicitSigmaPsi} where the point labeled ``$3$'' lies below the point labeled ``$1$''). At the critical point itself, one can see that there is no difference due to taking the mass of $\Psi$ into account, since in the quasi-stationary approximation, the result only depends on the instantaneous value of the mass, which vanishes precisely at the critical point. We will come back to this point in the next subsection. 

After the critical point, the calculations performed in the present work may be extrapolated for a few $e$-folds and one can see that the inclusion of the mass effects increases the value of $\sigma_\chi$, which makes sense since the fluctuation modes become tachyonic during the waterfall, hence the amplitude of the noise increases. However, when looking at the exact solution of Eq.~(\ref{eq:sigmapsiImplicit}), one can see that the actual value of $\sigma_\psi$ remains smaller. This can be understood as a time delay in the waterfall start. Indeed, when the fields system crosses the critical point, two minima in the $\Psi$ direction appear at 
\be 
\Psi_\pm^2=v^2\left(1-\frac{\Phi^2}{\Phi_\uc^2}\right)\, .
\ee
In between these two minima, the curvature of the potential in the $\Psi$ direction is negative whereas it is positive elsewhere. This is why when no back-reaction is taken into account, the fluctuations $\delta\psi^{(1)}$ become tachyonic as soon as the critical point is crossed. On the other hand, if back-reaction is ``switched on'' and if the fluctuations evolve about a $\sigma_\chi$-shifted background, the fluctuations keep on ``seeing'' a potential with positive curvature in the $\Psi$ direction as long as $\sigma_\psi>\left\vert\Psi_\pm\right\vert$. This means that the waterfall begins at a delayed time $N_\uc^\star$ instead of $N_\uc$, where $N_\uc^\star$ is defined by 
\be 
\label{eq:Ncstar}
\sigma_\chi\left(N_\uc^\star\right)=
\left\vert\Psi_\pm\left(N_\uc^\star\right)\right\vert=
v\sqrt{1-\frac{\Phi^2\left(N_\uc^\star\right)}{\Phi_\uc^2}}\, .
\ee
This ``effective'' critical time is displayed as the grey dotted line in Fig.~\ref{fig:SigmaPsi}. One can check that it coincides with the moment when the exact solution of Eq.~(\ref{eq:sigmapsiImplicit}) starts to strongly increase, \ie with the beginning of the ``effective'' waterfall phase. One could ask whether such an effect could save the model from the tachyonic breakdown of the perturbative expansion mentioned above. Indeed, if the waterfall start is sufficiently delayed so that it somehow ``never'' occurs, the effective potential curvature felt by the fields system is always positive and no pathological growth of the fluctuations occur. 

This can be rephrased as the following. Once the critical point crossed, the $\chi$-distribution splits into two pieces, each moving towards each minimum of the potential at $\Phi=0$, $\Psi=\pm v$. This is confirmed \eg by the numerical simulations of Ref.~\cite{Martin:2011ib} (see Fig.~10 there). Now, if one extends the quasi-stationary treatment presented above in the valley and assumes that the inflationary trajectory constantly tracks the local minimum in the $\chi$-direction, it implies that each piece of the distribution is centered over one of the two instantaneous minimums $\Psi_\pm\left(\Phi\right)$, so that most of the distribution settles over a positive potential curvature region. Obviously, this can occur only if the waterfall is sufficiently slowly driven by $\varphi$ so that a quasi-stationary distribution settles in the $\chi$-direction. This means that stochastic effects, combined with a long waterfall, may protect the hybrid model from the tachyonic issues mentioned above.

\subsection{Beyond the Quasi Stationary Approximation}

\begin{figure}
\begin{center}
\includegraphics[width=9cm]{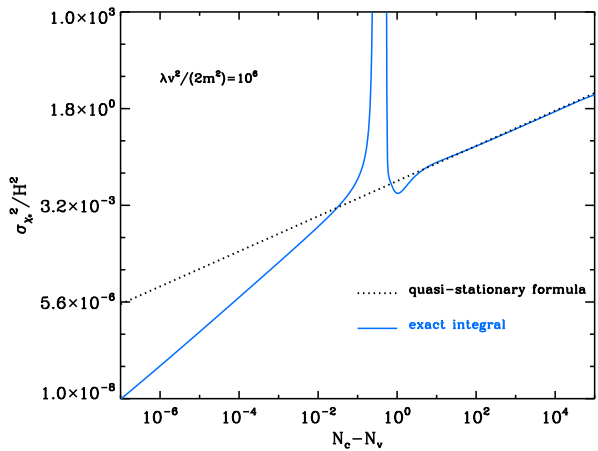}
\caption{Stochastic dispersion in the $\chi$ direction $\sigma_\chi^2$ at the critical point when $\varphi=\Phi_\uc$, normalized by the Hubble scale $H^2$, as a function of the number of $e$-folds spent between $N_{\mathrm{v}}$ and the critical point. The black dotted line corresponds to the massless formula~(\ref{eq:sigmapsic}), which is the expected result in the quasi-stationary approximation where the dispersion in the $\chi$ direction only depends on the instantaneous mass of $\chi$, which vanishes at the critical point. The blue solid line corresponds to the exact integral~(\ref{eq:sigmapsicExact}). The parameters are chosen as follows. If one defines $\alpha=\lambda v^2 /(2 m^2)$, the calculation can be shown to depend only on the two parameters $\alpha$ and $v/\Mp$. More precisely, the quantities appearing in Eqs.~(\ref{eq:sigmapsic}) and (\ref{eq:sigmapsicExact}) are $\alpha$, $\alpha v^2/\Mp^2$ [which roughly corresponds to the typical number of $e$-folds one can spend in the valley, see Eq.~(\ref{Nc})], and $\alpha v^4/\Mp^4$ [which both corresponds to the squared number of $e$-folds spent in the waterfall phase, see Eq.~(\ref{eq:efoldswater}), and to the number of $e$-folds spent between $N_{\mathrm{v}}$ and the critical point, see Eq.~(\ref{eq:Nv})]. In the figure, $\alpha$ is fixed to $\alpha=10^{6}$ and $v$ is varied below the Planck mass, and labeled by $N_\uc-N_{\mathrm{v}}$. One can check that the qualitative behaviour is independent on the chosen value for $\alpha$.}
\label{fig:sigmapsic}
\end{center}
\end{figure}

As already mentioned and as can be seen \eg in Eq.~(\ref{eq:efoldswater}), the number of $e$-folds realized in the waterfall phase depends on the typical dispersion in the $\chi$-direction at the critical point, $\sigma_{\chi_\uc}$. In the previous subsection, one has made use of a quasi-stationary approximation where $\sigma_\chi$ only depends on the instantaneous value of the $\Psi$-mass. At the critical point itself, this mass vanishes, hence no correction coming from the mass and its dynamical variation can be accounted for in this framework, and the obtained result coincides with the massless one~(\ref{eq:sigmapsic}). To check that this approximation scheme is reliable, and to identify the typical corrections appearing when it is not, the value of $\sigma_{\chi_\uc}$ is calculated in this section beyond the quasi-stationary approximation. Combining Eq.~(\ref{noisecorrpsi}) and Eq.~(\ref{eq:chistoSol}), one obtains the general formula
\bea 
\label{eq:sigmapsicExact}
\sigma_{\chi_\uc}^2&=&\frac{9}{8\pi^2}\frac{H^4}{m^2}
e^{\frac{\lambda v^2}{2 m^2}}
\quad\times\\& &
\int_1^\infty{e^{-\frac{\lambda v^2}{2m^2}\left(x-\ln x\right)}}
\left(\epsilon a H\right)^3 
\frac{\left\vert\delta\psi_{\bm{k}}^{(1)}\right\vert^2_{k=\epsilon a H}}{H^2}
\frac{\mathrm{d}x}{x}\, ,\nonumber
\eea
where one recalls that the argument of the integral is to be evaluated at $x\equiv \exp\left[-8\Mp^2m^2\left(N-N_\uc\right)/\left(\lambda v^4\right)\right]$. Making use of Eqs.~(\ref{PsimodesAnalBegin}-\ref{PsimodesAnalEnd}), this integral can be computed numerically. The result is displayed in Fig.~\ref{fig:sigmapsic}, as a function of the number of $e$-folds spent between $N_{\mathrm{v}}$ and the critical point. It is compared with the quasi-stationary formula, which coincides with the massless equation~(\ref{eq:sigmapsic}). The parameters are chosen according to what is explained in the caption of the figure. Three regimes of interest appear, that can easily be understood and described, keeping in mind the evolution of the effective mass $m_{\mathrm{u}}$ of the waterfall perturbations  displayed in Fig.~(\ref{fig:psimasssketch}).

When $N_\uc-N_{\mathrm{v}}\gg 1$, a large number of $e$-folds is spent between $N_{\mathrm{v}}$ and the critical point, which means that the effective mass of the waterfall perturbations varies slowly. In this case the quasi-stationary approximation is valid, the modes contributing the most to $\sigma_{\chi_\uc}^2$ are the ones that crossed their effective Hubble radius right before the critical point, which is far after $N_{\mathrm{v}}$. In Fig.~(\ref{fig:psimasssketch}), one can correspondingly check that the effective mass $m_{\mathrm{u}}$ is well approximated by the one of a massless field in this regime, and accordingly in Fig.~\ref{fig:sigmapsic}, the quasi-stationary formula and the exact integral match perfectly. One concludes that the quasi-stationary approximation holds for parameters such that $N_\uc-N_{\mathrm{v}}\gg 1$, which is equivalent to requiring a long lasting waterfall.

When $N_\uc-N_{\mathrm{v}}\ll 1$, a very small number of $e$-folds is spent between $N_{\mathrm{v}}$ and the critical point. Remembering that $m_{\mathrm{u}}$ vanishes at $N_{\mathrm{v}}$, this means that the effective mass of the waterfall perturbations varies very quickly and one expects the quasi-stationary approximation to break. More precisely, in this case $\sigma_{\chi_\uc}^2$ is still dominated by the modes that crossed their effective Hubble radius during, say, the last $e$-fold before turning the critical point, but because $N_{\mathrm{v}}\sim N_\uc$, they did so far before $N_{\mathrm{v}}$. In this regime the potential curvature in the $\Psi$-direction is not negligible anymore, and one can indeed check in Fig.~(\ref{fig:psimasssketch}) that the effective mass of the waterfall perturbations becomes much larger than the one for a massless field. The larger the potential curvature, the stronger it ``holds'' the field at its bottom, hence a decreased dispersion $\sigma_{\chi}$. This is exactly what is noticed in Fig.~\ref{fig:sigmapsic}, where for $N_\uc-N_{\mathrm{v}}\ll 1$, the dispersion is much smaller than what is predicted by the quasi-stationary formula.

Finally, these two cases are connected by the regime $N_\uc-N_{\mathrm{v}}\sim 1$ where a more singular behavior occurs. In this case, $\sigma_{\chi_\uc}^2$ is again dominated by the modes that crossed their effective Hubble radius during the last few $e$-folds before turning the critical point, that is exactly around $N_{\mathrm{v}}$ since $N_\uc-N_{\mathrm{v}}\sim 1$. Remembering that the effective mass of the waterfall perturbations vanishes at $N_{\mathrm{v}}$, one can check in Fig.~(\ref{fig:psimasssketch}) that there is a small time interval around $N_{\mathrm{v}}$ during which $m_{\mathrm{u}}$ is suppressed, and much smaller than its massless counterpart. During this short period $\chi$ diffuses almost freely, hence the peak noted in Fig.~\ref{fig:sigmapsic}. This regime is, however, rather fine tuned, since there is \apriori no reason why $N_\uc-N_{\mathrm{v}}\sim 1$.

In conclusion, the quasi-stationary approximation which enables to develop the calculations and the results of the previous subsection holds in the regime of parameters for which a long waterfall occurs (or equivalently $N_\uc-N_{\mathrm{v}}\gg 1$), and when it is not the case, the actual dispersion in the $\chi$-direction is decreased. However, since the number of $e$-folds spent in the waterfall precisely depends on this typical dispersion at the critical point, one can see that even in this regime, this number of $e$-folds may not be that small.

\section{Inflaton perturbations beyond zeroth order in slow-roll}
\label{sec:influcts}

Let us now recapitulate our progress so far. The formalism of stochastic inflation has allowed us to split the full quantum system formed by the two scalar fields in quasi-de Sitter space into a quantum bath and a coarse-grained, classical system, and to integrate out the bath to obtain an effective theory for the coarse-grained system. In this effective theory formalism, the quantum effects are modeled through a classical stochastic noise term in the equation of motion for each field, which can be calculated from the propagator of the quantum fields.

Assuming propagators valid up to zeroth order in slow-roll, we have obtained a first approximation for the coarse-grained fields dynamics, valid to $\mathcal{O}(\hbar)$ and zeroth order in slow-roll in section~\ref{sec:ClassAndModes}. However, many cosmological observables are known to depend primarily on higher order quantities, for example the spectral index of curvature perturbations, $n_s$. To increase the order of precision of our answer, we had to calculate the noise sourcing the Langevin equations to higher order. This is what we did in section~\ref{sec:PertExpAndModeCalc}, where we calculated the linearized mode functions for the quantum fields in the presence of a stochastically shifted background, which allowed us to obtain shifted noise amplitudes valid to leading order in slow-roll and up to $\mathcal{O}(\hbar^2)$. Note that it did not allow us, however, to calculate the corrected classical observables, such as the spectral index, because in the stochastic formalism these are quantities which must be calculated from perturbations of the classical stochastic system, rather than the quantum bath.

We then moved on to calculate the classical effects of a shifted noises on the one-point statistics of the coarse-grained waterfall field in section~\ref{sec:sigmapsi}, insisting in particular on its dispersion as the critical point is approached. Now that we have developed a good understanding of the coarse-grained waterfall field behaviour beyond $\mathcal{O}(\hbar, \varepsilon_1^0, \varepsilon_2^0)$, we can turn to the question of how stochastic effects will affect the details of the statistics of the coarse-grained inflaton field to leading order in slow-roll (and to $\mathcal{O}(\hbar^2)$). In particular, we will be interested in calculating how stochastic effects modify the tilt of the curvature perturbations power spectrum.

First, we need to incorporate the shifted noise from section~\ref{sec:PertExpAndModeCalc} in the Langevin equation. This is the noise amplitude we already wrote in Eq.~(\ref{theshiftedinflatonnoisespectrum}) and which we re-write here for clarity:
\bea 
\label{eq:xiphiCorr}
\left\langle\xi_\phi\left(N\right)\xi_\phi\left(N^\prime\right)\right\rangle=
\frac{H^4}{4\pi^2}\delta\left(N-N^\prime\right)
\times\nonumber\\
\left[1+\frac{2}{3}\left(\frac{m^2+g\sigma_\chi^2}{H^2}\right) 
\left(\ln 2\epsilon +\gamma-2\right)\right]\, .
\eea
Note that, in the following, we will only keep the leading contribution from the second slow-roll parameter $\varepsilon_2$, since we want to capture the leading effect in magnitude and $\varepsilon_2^2\gg\varepsilon_1$ for the values of parameters we are considering. Since we are neglecting all powers of $\varepsilon_2$ higher than one, we have neglected the factor of $\varepsilon_1$ in the derivation presented bellow.

From there, to address the question of the classical coarse-grained inflaton spectrum, one would technically need to solve the Fokker-Planck equations corresponding to Eqs~(\ref{Langphi})-(\ref{Langpsi}) with the noises calculated from the results of section~\ref{sec:PertExpAndModeCalc} through equations~(\ref{noisecorrphi})-(\ref{noisecorrpsi}). However, this turns out to be a rather difficult task analytically, and the result not readily useable to get concrete observable predictions.

Fortunately, we can perform a simpler calculation which circumvents the difficulties of solving the Fokker-Planck equations. From the previous section, we already obtained a solution of Eq.~(\ref{Langpsi}) to derive the mean and dispersion of $\chi$, which holds provided that $\langle \varphi\rangle$ remains close to the classical, noiseless solution (we have verified this is indeed the case for the regime of parameters we are considering in the current work, {\it i.e.} $\Delta_\phi\ll1$). We could perform a similar analysis for $\varphi$, but this would not be of much help since we are really interested in separating the power in $\varphi$ coming from the ``mean'' uniform background {\it classical} evolution, and the one coming from the fluctuations in $\varphi$ which give rise to the power spectrum in the CMB. 

The strategy we adopt is therefore to expand Eq.~(\ref{Langphi}) as follows\footnote{Note, however, that even though we are splitting the classical fields into `mean' classical field and classical perturbations, the mean background felt by the quantum fields is still $\langle \varphi=\varphi_0+\delta\varphi^{(1)}+...\rangle$, and similarly for higher powers.}:
\be 
\label{linearclass} \varphi=\varphi_0+\delta\varphi^{(1)}+... .
\ee

Our goal here is to find the average power in the linear inflaton classical fluctuations squared $\langle (\delta\varphi^{(1)})^2\rangle$, analogously to what is done in Ref.~\cite{Finelli:2008zg}, and then take its time-derivative to recover the $k$-dependence of its power spectrum. Using that the noises should be treated perturbatively, we obtain the usual $\frac{\dd\varphi_0}{N}=-\frac{-V_\phi}{3H^2}$ for the equation of motion of the classical mean $\varphi_0$ field, while for the linear perturbations $\delta\varphi^{(1)}$ we obtain: 
\be 
	\frac{\dd\delta\varphi^{(1)}}{\dd N} +2\Mp^2\left(\frac{H_{,\Phi}}{H}\right)_{,\Phi}\delta\varphi^{(1)} =\frac{\xi_\phi}{H} \, ,
\ee
where, as before, $\xi_\phi$ is the contribution of the stochastic noise in $\varphi$. Here, the occurrences of $\chi$ in $H$ are the full coarse-grained fields since we are not doing an expansion in the coarse-grained waterfall field, only in the coarse-grained inflaton field.
Multiplying this equation by $\delta\varphi^{(1)}$ and taking the average, we obtain:
\bea
\label{eq:deltavarphi1}
\frac{\dd \langle (\delta\varphi^{(1)})^2\rangle}{\dd N}+4\Mp^2\left(\frac{H_{,\Phi}}{H}\right)_{,\Phi}\langle (\delta\varphi^{(1)})^2\rangle = \nonumber\\ \frac{H^2}{4\pi^2}\left(1+\frac{2}{3}\frac{A}{H^2} \right)\, ,
\eea
where we have defined $A=\tilde{m}^2(\ln 2\epsilon +\gamma -2)$, with $\tilde{m}^2=(m^2+g^2\sigma_\chi^2)$, and where we assumed $\delta \varphi^{(1)}$ and $\chi$ are mutually independent and used $\langle\chi\rangle=0$, as well as $\langle\chi^2\rangle=\sigma_\chi^2$, which have already been calculated at the required order. We have also used the relation\footnote{This relation can be obtained plugging a formal solution of Eq.~(\ref{eq:deltavarphi1}), $\delta\varphi^{(1)}=\int\dd N\left[f/H-2\Mp^2(H_{,\Phi}/H)_{,\Phi}\delta\varphi^{(1)}\right]$, into $\langle \xi_\phi \delta\varphi^{(1)} \rangle$, and using Eq.~(\ref{eq:xiphiCorr}) as well as the identity $\int_a^{x_0}\delta(x-x_0)f(x)dx=f(x_0)/2$.} $\langle \xi_\phi \delta\varphi^{(1)} \rangle =H^3\left(1+\frac{2}{3}\frac{A}{H^2} \right)/(8\pi^2)$.

Integrating and using the zeroth order equation to re-write the solution in terms of an integral over $\varphi$, we get the solution:
\bea
	\langle (\delta\varphi^{(1)})^2\rangle=\qquad\qquad\qquad\qquad\qquad\qquad\qquad\qquad\qquad\\
	\nonumber\left(\frac{H_{,\Phi}}{H}\right)^2\frac{1}{8\Mp^2}\int_{\varphi_0}^{\varphi_{0,\uin}}\left( \frac{H^5}{H_{,\Phi}^3}\right)\left(1+\frac{2}{3}\frac{A}{H^2} \right)\dd \varphi\, .
\eea
Using the solutions for $\varphi_0$ and $H_0$, this integral can easily be performed keeping expressions for $H$ to leading order in $m^2$. We obtain (by analogy to \eg Ref.~\cite{Finelli:2010sh}):
\bea
	\langle (\delta\varphi^{(1)})^2\rangle\approx \frac{3 H^4\varphi_0^2}{8\pi^2 \tilde{m}^2}\left[1-\frac{\varphi^2_0}{(\varphi_0)_\uin^2} \right]\left(1+\frac{2}{3}\frac{A}{H^2} \right)\, .
\eea
This result is sensible since at the beginning of inflation, when $\varphi^2_0=(\varphi_0)_\uin^2$, there is no power in the inflaton fluctuations. As inflation proceeds and the classical background inflaton rolls downs its potential, there is more and more power (qualitatively because modes are joining the coarse-grained field, and doing so adding power to the classical fluctuations) and at sufficiently late times the system approaches a ``quasi-equilibrium'' average power in the fluctuations\footnote{This picture holds given our assumption that $H$ is truly constant. In a more realistic scenario, this is only approximately true but can still provide intuition on what is actually happening.}. If were to carry through and calculate the tilt induced by this piece of the time-dependence of $\langle (\delta\varphi^{(1)})^2\rangle$, we would obtain a contribution to the final tilt which is subdominant\footnote{More specifically, its contribution to the tilt is blue, but initially less by a half than the contribution to the tilt we calculate in what follows, and it has a decaying pre-factor which becomes negligible as this ``quasi-equilibrium'' is approached.}. We therefore neglect the time-dependence coming from $\varphi^2_0/(\varphi_0)_\uin^2$ in the remaining of this calculation.

Comparing with the usual QFT methods, we know that the general formula for massive modes far outside the Hubble radius is given by~\cite{Finelli:2003bp}:
\bea
	\phi_k=\frac{1}{a^{3/2}}\left( \frac{\pi \lambda}{4H}\right)^{1/2}\left[ \frac{H(t_k)}{H(t)}\right]^2H^{(1)}_{3/2}\left[\frac{k(1+\varepsilon)}{\epsilon aH}\right] \, ,\nonumber\\ \\
	\mathrm{with} \ \ H(t_k)=H_\uin\sqrt{1+2\frac{\dot H_\uin}{H_\uin^2}\ln\left[ \frac{(1+\varepsilon_\uin)k}{H_\uin\nu_\uin}\right]}\,.\nonumber\\
\eea
Therefore, when one is interested in the average power in the fluctuations, one needs to calculate the following integral:
\bea
	\langle \phi_k^2 \rangle^{\mathrm{IR}}= \frac{1}{4\pi^2}\left( \frac{H_\uin}{H}\right)^2\frac{H_\uin^2}{(1+\varepsilon)^2}  \int_{l}^{\epsilon aH} D^2(k)\, ,
\eea
where we have defined the function $D(k)$ to have only $k/aH$ and $\nu$ dependence and no other time dependence (all the modes' time dependence has been brought to the front of the integral).

Therefore, we find that:
\bea
	\int_ l^{\epsilon aH}\frac{\dd k}{k} k^3\left|\delta \varphi^{(1)}_k\right|^2 \qquad \qquad\qquad \qquad\qquad \qquad\qquad~  \\\sim 4\pi^2(1+\epsilon)^2\frac{H^6}{H_\uin^4}\frac{\varphi^2_0}{\tilde{m}^2}\frac{3}{8\pi^2}\left[1-\frac{\varphi^2_0}{(\varphi_0)_\uin^2} \right]\left(1+\frac{2}{3}\frac{A}{H^2} \right)\nonumber\\
	\approx \frac{3H^4}{2\tilde m^2}\left(1+\frac{2}{3}\frac{A}{H^2} \right)\left[ 1-\frac{\varphi^2_0}{(\varphi_0)_\uin^2} \right] \qquad\qquad \qquad\qquad 
\eea
where in the last line we have used that at this order in $m$, $H$ is a constant.
From the leading coefficient, we recognize the standard result for the blue-tilted spectrum of a massive field. We therefore obtain:
\bea
	\frac{\dd k}{k} k^3\left|\delta \varphi^{(1)}_k\right|^2 &\propto&\frac{\dd k}{k} \left(\frac{k}{aH} \right)^{\frac{2\tilde m^2}{3H^2}\frac{1}{\left(1+\frac{2}{3}\frac{A}{H^2}\right)}}\\
	\Rightarrow  k^3\left|\delta \varphi^{(1)}_k\right|^2&\approx&  \left(\frac{k}{aH} \right)^{\frac{2\tilde m^2}{3H^2}-\frac{4}{9}\frac{\tilde m^2}{H^2}\left(\frac{\tilde m^2}{H^2}\right)(\ln 2\epsilon +\gamma -2)}
\eea

Here the second term in the exponent is the one coming from the modified amplitude of the noise sourcing the $\delta \varphi^{(1)}$ equation of motion, while the fact that we took the full $\chi$ field to source the mass of $\varphi_0$ is the reason why $\tilde m^2$ appears instead of the usual $m^2$.  Even though, at this order, the conceptually different methods of, on one side, perturbing the classical coarse-grained inflaton to obtain its classical spectrum and, on the other side, reading it off from the spectrum of quantum mode functions directly give the same result, there is no guarantee that this will indeed be the case when one computes higher order corrections in slow roll. One should therefore be careful when it comes to taking short cuts to obtain observables in stochastic inflation, as this expansion strategy separates the bath and the coarse-grained system into distinct theories sourcing each other.

Note that the shifted noise has a contribution which is higher order in $\tilde m^2/H^2$, in such a way that it actually gives rise to a correction which is higher order in $\hbar$. Thus, we cannot retain it whilst neglecting contributions of similar order coming from different sources.
Hence, we find as our final result that the spectral index to leading order in slow-roll is:
\be
	\nS=1+\frac{2(m^2+g^2\sigma_\chi^2)}{3H^2}= 1+\frac{2g^2\sigma_\chi^2}{3H^2}-\varepsilon_2\, .
\ee 
This result (which is the main result of this section) can be understood as being the standard one provided that one performs the replacement $m^2\rightarrow \tilde m^2$ for the mass of the inflaton, which comes from using the shifted $\chi$ rather than the zeroth order background value $\chi^ {(0)}=0$. The interesting point here is that this modification of the standard spectral index formula shows an example of resummed quantum corrections competing with the usual slow-roll corrections. Indeed, since $g^2\sigma_\chi^2$ can be comparable to $m^2$, mainly close to the critical point, there is a region of parameter space where stochastic corrections can dominate over slow-roll effects. 

Finally, and more importantly, since $\tilde m^2 > m^2$, the stochastic dispersion of $\chi$ makes the inflaton more massive. Therefore, as suspected by looking at the spectrum of the quantum fluctuations causing the noise, the spectrum becomes bluer due to stochastic effects. Moreover, we obtain that the tilt is modified by an $\mathcal{O}(1)$ factor compared to an estimate based solely on slow-roll parameters. This is one of the main results of the paper. 

Note that this effect is however not expected to occur in all models of inflation, since it is due to the particular way various mass scales are set in hybrid inflation. In particular, the reason why metric perturbations cannot overcome the tendency of the mass of the inflaton $m$ to make the tilt blue is because the first slow-roll parameter is set by the vacuum energy dominating $H$, which is independent of the adiabatic direction in the potential. In other words, the ratio $m^2/H^2$ is proportional to the second slow-roll parameter, rather than the first as is the case in single field inflation. As the system approaches the critical point, the waterfall field becomes lighter, and its dispersion approaches that of a light field, {\it i.e.} becomes comparable to that of the inflaton, allowing the two corrections to the tilt to be comparable in size if the transition is sufficiently slow.

\section{Conclusion}
\label{sec:conclusion}

In this paper we have investigated the effects of a recursive stochastic approach to the valley phase of hybrid inflation, making use of the method presented in Ref.~\cite{paper1}, where the noise amplitude is calculated from the scalar perturbations evolving about a background continuously shifted by the modes sourcing the coarse-grained fields.

This paper therefore presented an illustration of how to implement consistently this recursive method of stochastic inflation in multi-field cases, and applied it to derive novel interesting results. In particular, it provided a concrete example where leading corrections to observables can be dominated by stochastic effects rather than slow-roll parameters. In the valley of the hybrid potential, it was found that this consistent calculation yields a blue tilt problem which is worse by an $\mathcal{O}(1)$ factor compared with the usual slow roll contribution. This indicates that if one wishes to modify the valley potential to generate a red tilt, it is crucial to take into account the stochastic contribution to the spectral index.

It was also demonstrated how to obtain the correct dispersions at a given order for both the inflaton and the waterfall fields. The latter sets the length of the waterfall, which in turn can potentially determine the viability of the model, and must therefore be computed accurately.  Short-lived waterfalls were shown to be unlikely, since the quasi-stationary time behavior of the auxiliary field distribution breaks down in this regime, reducing its quantum dispersion at the critical point, hence lengthening this final stage. Besides, short-lived waterfalls imply that the long wavelengths of the auxiliary field do not experience quantum squeezing, in which case the usual interpretation of the stochastic formalism is problematic. Furthermore, an analysis of back-reaction showed that the recursive process converges in the valley but fails during the waterfall, suggesting the presence of an expected perturbative instability. 

Even though to find a regime where the spectral tilt $n_{\mathrm{S}}$ is compatible with current constraints a long waterfall phase containing the observational window may seem like an attractive solution, the tachyonic growth of the waterfall field and the exponential growth of entropy scalar perturbations make a traditional perturbative approach unstable and out of control in this final stage. If at all, a solution may be found if the stochastic effects combined with a long and slow waterfall phase allow for the fields distribution to continuously settle over the two local $\Psi$-minimums in a quasi-stationary way. This is why it becomes crucial to be able to consistently compute the physical predictions of such a genuine two field phase, properly including the stochastic contribution on the background. This shall be the purpose of future work.

\acknowledgments

The authors would like to thank Giovanni Marozzi for useful discussions.
The work of LPL is supported in part by an NSERC PGS-D and an M.T. Meyers scholarship from Girton College. The research of RB is supported in part by a NSERC Discovery Grant and by funds from the Canada Research Chair program.

\appendix

\section{Notations and Assumptions on the Parameters}
\label{app:NotationsAndAssumptions}

In this appendix we summarize the notations used in this paper, as well as the assumptions made on the potential parameters. The potential of hybrid inflation is given by
\be
	V(\Phi,\Psi)=\frac{1}{2}m^2\Phi^2+\frac{\lambda}{4}(\Psi^2-v^2)^2+\frac{g^2}{2}\Phi^2\Psi^2 \, , \nonumber
\ee
where $\Phi$ and $\Psi$ are the inflaton and waterfall fields, $g$ and $\lambda$ are supposedly small coupling constants, $m$ is the mass of the inflaton, and $v$ is the \vev of the waterfall at the global minima of the potential $\Phi=0$, $\Psi=\pm v$. The critical point is located at $\Phi=\Phi_\uc\equiv v\sqrt{\lambda}/g$, $\Psi=0$, and the ``valley'' corresponds to $\Phi>\Phi_\uc$, $\Psi\simeq 0$. If the model is derived in the framework of supersymmetry, one has
\be 
\Phi_\uc=v \Rightarrow \lambda=g^2\, .
\ee
For inflation to proceed at ``small field'' values, these parameters $\Phi_\uc$ and $v$ must be small compared to the Planck mass
\be 
\Phi_\uc,v\ll \Mp\, .
\ee 
The vacuum dominated regime corresponds to \vevs of the fields for which the potential is dominated by its constant term $V\simeq\lambda v^4/4$, that is $\Psi\ll v$, and $\Phi_\uc<\Phi\ll \lambda v^2/m$. The former is well verified in the valley, even if one starts from sizable values of $\Psi_\uin/v$ (in which case the bottom of the valley is reached very quickly), and even in the presence of stochastic effects, as shown \eg after Eq.~(\ref{eq:sigmapsic}), while the later implies that
\be 
gv\gg m\, .
\ee
It is also assumed that a slow roll regime of inflation takes place in the valley. The smallness of the first slow roll parameter $\varepsilon_1\ll 1$ implies that
\be 
\lambda v^4\gg m^2\Phi_\uc\Mp\, ,
\ee
while the smallness of the second slow roll parameter $\varepsilon_2\ll 1$ implies the more stringent condition
\be 
\lambda v^4\gg m^2\Mp^2\, .
\ee 
Finally, to avoid the blue tilt problem one may wish to realize the last $\sim 60$ $e$-folds of inflation in the waterfall stage. From Eq.~(\ref{eq:efoldswater}) this is the case only if
\be 
\sqrt{\lambda}v^3\gg m\Mp^2\, .
\ee 

We now explain the notation employed to refer to different quantities associated with each quantum field. In Eq.~(\ref{hybridInflationpotential}), the potential was written in terms of the full quantum operator fields $\Phi$ and $\Psi$. Their classical homogeneous background counterpart are denoted by $\varphi^{\left(0\right)}$ and $\chi^{\left(0\right)}$. $\Phi$ and $\Psi$ are Fourier expanded in terms of the classical mode functions $\phi_{\bf k}$ and $\psi_{\bf k}$ (and the creation and annihilation operators $\hat{a}^\dagger_{\bf k}, \hat{a}_{\bf k}, \hat{b}^\dagger_{\bf k}, \hat{b}_{\bf k}$).

One collects the small wavelength modes of the full quantum fields to define the quantum bath $\phi_>$ and $\psi_>$, with their linearized counterparts denoted by $\delta\phi_>^{\left(1\right)}$ and $\delta\psi_>^{\left(1\right)}$. The large wavelength-modes collectively form the classical stochastic coarse-grained system fields $\varphi$ and $\chi$, formally defined by $\varphi=\Phi-\phi_>$ and $\chi=\Psi-\psi_>$. Classical linearized fluctuations around the coarse-grained fields are denoted $\delta\varphi^{(1)}$ and $\delta\chi^{(1)}$.

\section{Classical Dynamics of the Waterfall Phase}

\label{app:waterfall}

Following the terminology used in Ref.~\cite{Kodama:2011vs}, this phase can be divided into three consecutive sub-phases.

``Phase-$0$'' consists in neglecting the last term in the inflaton slow-roll equation~(\ref{KGphi}) and the first one on the right hand side of the waterfall equation~(\ref{KGpsi}) (on the ground that, initially $\varphi=\Phi_\mathrm{c}$). The slow-roll solutions read
\begin{eqnarray}
\label{eq:solphi0}
\varphi^{\left(0\right)}(N) &=& \Phi_\mathrm{c} 
\exp\left[-4 \frac{\Mp^2m^2}{\lambda v^4}
\left(N-N_\mathrm{c}\right)\right]\, ,\\
\label{eq:solpsi0}
\chi^{\left(0\right)}(N) &=& \chi _\mathrm{c} 
\left[1+\frac{8\Mp^2\chi_\mathrm{c}^2}{v^4}
\left(N-N_\mathrm{c}\right)\right]^{-1/2}\, ,
\end{eqnarray} 
where $N_\mathrm{c}$ denotes the number of $e$-folds at the critical point, \ie at the onset of the waterfall phase. This phase ends when $\varphi=\varphi_1$ and $\chi=\chi_1$, with
\begin{eqnarray}
\ln \frac{\varphi_1}{\Phi_\mathrm{c}}&\simeq &
\frac{m^2}{4\lambda \chi_\mathrm{c}^2}
\left(1-\sqrt{1+\frac{4\lambda \chi_\mathrm{c}^4}{m^2v^2}}\right)\, ,
\nonumber\\
\chi_1&\simeq &v\sqrt{-2\ln \frac{\varphi_1}{\Phi_\mathrm{c}}}\, .
\end{eqnarray}
If we are in the regime where $4\lambda \chi_\mathrm{c}^4/(m^2v^2)\ll 1$, then the number of $e$-folds realized in this phase is given by
\begin{equation}
N_1-N_\mathrm{c}\simeq \frac{\lambda v^2\chi_\mathrm{c}^2}{8\Mp^2m^2}\ll 1,
\end{equation}
where $N_1$ denotes the number of $e$-folds at the end of phase-$0$. In practice, $\chi_\mathrm{c}/v$ is so small that $N_1-N_\mathrm{c}$ is always very small. In this case, we conclude that the phase-$0$ is unimportant since it lasts a negligible number of $e$-folds and since the values of $\varphi$ and $\chi$ remain almost unchanged during that phase.

We now proceed with Phase-$1$, where the second term on the right hand side of the waterfall equation~(\ref{KGpsi}) can be neglected. During this phase, the solution for the inflaton field is unchanged, but the waterfall field
evolution now reads 
\begin{equation}
\label{eq:psi0Waterfall1}
\chi^{\left(0\right)}=\chi_1\exp\left\{\frac{16m^2\Mp^4}{\lambda v^6}
\left[\left(N-N_\mathrm{c}\right)^2 -
\left(N_1-N_\mathrm{c}\right)^2\right]\right\}.
\end{equation}
The Phase $1$ stops when the first term on the right hand side of the waterfall
field equation becomes important, {\it i.e.} when $\chi\equiv \chi_2$ and $\varphi\equiv\varphi_2$, where
\begin{eqnarray}
\label{eq:psi2}
\chi_2^2&=&\frac{\Phi_\mathrm{c}^2m^2}{\lambda v^2}=\frac{m^2}{g^2}\, ,\\
\ln ^2\frac{\varphi_2}{\Phi_\mathrm{c}}&\simeq & \frac{m^2}{\lambda v^2}\ln 
\left(\frac{m}{g\chi_\mathrm{c}}\right)\, .
\label{eq:phi2}
\end{eqnarray}
Finally, the number of $e$-folds produced during Phase-$1$ is given by
\begin{equation}
\label{eq:N2}
N_2-N_\mathrm{c}\simeq \frac{\lambda^{1/2}v^3}{4m\Mp^2}\ln ^{1/2}\left(
\frac{m}{g \chi_\mathrm{c}}\right)\, .
\end{equation}
Therefore, if one were interested in the regime where the required $60$ $e$-folds of inflation take place during the waterfall phase, one needs to work in the $\frac{\lambda v^6}{m^2\Mp^4}\gg 1$ regime.

Finally, let us now briefly mention Phase-$2$, where one needs to keep the last term in the inflaton equation of motion~(\ref{KGphi}), hence equations~(\ref{KGphi}) and~(\ref{KGpsi}) become fully coupled. The slow-roll trajectory in field space obeys
\begin{equation}
\chi^2=\chi_2^2+\varphi^2-\varphi_2^2-2\Phi_\mathrm{c}^2
\ln \frac{\varphi}{\varphi_2}\, .
\end{equation}
During Phase-$2$, inflation quickly stops and the system starts oscillating around one of the two true minimums of the potential.

\section{Formulas for $\delta\psi_{\bm{k}}^{(1)}$}
\label{app:deltapsiFormula}

In this appendix we sumarize, for practical convenience, the derived formula for the amplitude of the first order perturbations in the 
$\Psi$ direction $\delta\psi_{\bm{k}}^{(1)}$. Defining
\bea
&x\left(N\right)=\left[\frac{v^2H^2}{8m^2\Mp^2}\right]^{\frac{2}{3}}
\left[\frac{8m^2\Mp^2}{v^2H^2}(N-N_c)-3
\frac{\lambda\sigma_\chi^2}{H^2}+15\varepsilon_1+\frac{9}{4} \right] 
\, ,&\nonumber\\
\eea
one has
\begin{widetext}
\bea
\label{PsimodesAnalBegin}
\mathrm{If }& & k<k_\mathrm{v}=\frac{Hv}
{\sqrt{6}\Mp}\mathrm{e}^{N_\uc-\frac{v^2H^2}{4\Mp^2m^2}}\, ,\nonumber\\
& &\quad\quad\mathrm{if}\quad N<N_{H_\mathrm{eff}}\simeq
\frac{1}{2}\log\left(\frac{v^2}{\Mp^2}\frac{k^2}{12H^2}\right)-8
\frac{\Mp^2m^2}{\lambda v^4}N_\uc\nonumber\, ,\\
& &\quad\quad\quad\quad\quad\quad
\left\vert\delta\psi_{\bm{k}}^{(1)}\right\vert\approx
\frac{\mathrm{e}^{-N}}{\sqrt{2k}}\, ,\\
& &\quad\quad\mathrm{if}\quad 
N_{H_\mathrm{eff}}<N<N_{\mathrm{v}}=N_\uc-\frac{v^2H^2}{4\Mp^2m^2}\nonumber\, ,\\
& &\quad\quad\quad\quad\quad\quad
\left\vert\delta\psi_{\bm{k}}^{(1)}\right\vert\approx
\mathrm{e}^{\frac{1}{2}\left[N_{H_\mathrm{eff}}-3N\right]}
\left\vert\frac{x(N_{H_\mathrm{eff}})}{x(N)}\right\vert^{\frac{1}{4}}
\frac{1}{\sqrt{2k}}\, ,\\
& &\quad\quad\mathrm{if}\quad N_{\mathrm{v}}<N<N_\mathrm{c}\nonumber\, ,\\
& &\quad\quad\quad\quad\quad\quad
\left\vert
\delta\psi_{\bm{k}}^{(1)}\right\vert\approx
e^{\frac{1}{2}(N_{H_\mathrm{eff}}-3N)}\left\vert\frac{x(N_{H_\mathrm{eff}})}
{x(N)}\right\vert^{\frac{1}{4}}\frac{e^{\frac{2}{3}x(N)^{\frac{3}{2}}
-\frac{2}{3}x(N_\mathrm{v})^{\frac{3}{2}}}}{\sqrt{2k}}\, ,
\eea

\bea
\mathrm{If }& 
&k_\mathrm{v}<k<k_{\mathrm{c}}=\sqrt{2}H\mathrm{e}^{N_\mathrm{c}}\nonumber\, ,\\
& &\quad\quad\mathrm{if}\quad 
N<N_{H_\mathrm{eff}}\simeq\left[\log\left(\frac{k}
{\sqrt{2}H}\right)+\frac{24m^2\Mp^4}{\lambda v^6}\frac{k^2}
{H^2}N_\mathrm{c}\right]\Big{/}\left(1-\frac{24m^2\Mp^4}
{\lambda v^6}\frac{k^2}
{H^2}\right)\nonumber\, ,\\
& &\quad\quad\quad\quad\quad\quad
\left\vert\delta\psi_{\bm{k}}^{(1)}\right\vert\approx\frac{\mathrm{e}^{-N}}
{\sqrt{2k}}\, ,\\
& &\quad\quad\mathrm{if}\quad N_{H_\mathrm{eff}}<N<N_\mathrm{c}\nonumber\, ,\\
& &\quad\quad\quad\quad\quad\quad
\left\vert\delta\psi_{\bm{k}}^{(1)}\right\vert\approx
\mathrm{e}^{\frac{1}{2}\left[N_{H_\mathrm{eff}}-3N\right]}
\left\vert\frac{x({N_{H_\mathrm{eff}}})}{x(N)}\right\vert^{\frac{1}{4}}
\mathrm{e}^{\frac{2}{3}\left[x\left(N\right)^{\frac32}-
x\left(N_{H_\mathrm{eff}}\right)^{\frac32}\right]}
\frac{1}{\sqrt{2k}}\, .
\label{PsimodesAnalEnd}
\eea
\end{widetext}

\pagebreak
\null
\newpage 

\bibliography{StochasticRecursiveHybrid.bib}

\end{document}